\documentclass[aps,reprint,twocolumn,showpacs,floatfix,superscriptaddress]{revtex4-2}
\usepackage{amsmath,amssymb,color,braket,hyphenat,makeidx,xcolor}
\usepackage[ocgcolorlinks,colorlinks=true,linkcolor=blue,citecolor=red,linktocpage=true]{hyperref}

\usepackage{graphicx}
\usepackage[caption=false]{subfig}

\begin{document}
	\title{Correlation Enhanced Autonomous Quantum Battery Charging via Structured Reservoirs}
	\author{A. Khoudiri}
	\email{khoudiri.achraf@etu.uae.ac.ma}
	\affiliation{Laboratory of R$\&$D in Engineering Sciences, Faculty of Sciences and Techniques Al-Hoceima, Abdelmalek Essaadi University, Tetouan, Morocco.}
	 \affiliation{The UM6P Vanguard Center, Mohammed VI Polytechnic University (UM6P),\\
		Rocade Rabat-Salé, Technopolis,11103, Morocco.}
	\author{A. El Allati}
	\email{eabderrahim@uae.ac.ma}
	\affiliation{Laboratory of R$\&$D in Engineering Sciences, Faculty of Sciences and Techniques Al-Hoceima, Abdelmalek Essaadi University, Tetouan,	Morocco.}
	\author{Y. Khlifi}
	\affiliation{Laboratory of R$\&$D in Engineering Sciences, Faculty of Sciences and Techniques Al-Hoceima, Abdelmalek Essaadi University, Tetouan,	Morocco.}
	\affiliation{IGFAE, University of Santiago de Compostela, E-15782, Santiago de Compostela, Spain.}
	\author{K. El Anouz}
	\email{kelanouz@uae.ac.ma}
	\affiliation{Laboratory of R$\&$D in Engineering Sciences, Faculty of Sciences and Techniques Al-Hoceima, Abdelmalek Essaadi University, Tetouan,	Morocco.}
	\author{Ö. E. Müstecaplıoğlu}
	\email{omustecap@ku.edu.tr}
	\affiliation{Department of Physics, Koç University, Istanbul, Sarıyer 34450, Türkiye.}
	\affiliation{TÜBİTAK Research Institute for Fundamental Sciences, Gebze, 41470, Türkiye.}
	
	\begin{abstract}
		In this work, we investigate the autonomous charging process of a quantum
			battery coupled to a structured reservoir composed of two qubits, each locally coupled to its own bosonic thermal bath. Moreover, the reservoir interacts with a charger–battery architecture through three configurations: (I) direct coupling between reservoir qubits and battery, (II) collective coupling among the reservoir qubits, charger, and battery, while (III) reflects a collective coupling between the reservoir qubits and charger together with a local charger–battery interaction. However, by using incoherent and coherent initial states, we analyze the stored energy, ergotropy, and charging power of the battery, and derive upper and lower bounds on the extractable work in terms of the free energy of coherence and the correlations exchanged between subsystems. Our results show that global and local coherences, as well as total correlations, act as quantum resources that enhance autonomous charging. Additionally, we demonstrate that the free energy stored in the quantum battery splits into contributions from coherence and correlations, providing numerical evidence that supports the derived ergotropy bounds. Importantly, this work highlights how structured reservoirs enable autonomous and resource-enhanced quantum battery operation.

	\end{abstract}
	\pacs{
		05.70.Ln,  
		05.30.-d   
		03.67.-a   
		42.50.Dv   
	}
	\maketitle
	
	\section{Introduction}
	
In recent years, the world has shown increasing interest in quantum technology applications~\cite{intro1,intro2}. Quantum batteries are among the most promising applications in the context of quantum thermodynamics~\cite{intro3,intro4}. They have been studied in various scenarios. For instance, Farina \textit{et al.} presented a model based on coherence-driven quantum batteries in contact with a thermal reservoir, exploring several battery and charger configurations~\cite{intro5}. Moreover, Andolina \textit{et al.} analyzed different classifications and combinations of quantum harmonic oscillator batteries in the context of charger-mediated energy transfer~\cite{intro6}. More recently, Ref.~\cite{intro6a} investigated the physical mechanisms governing ergotropy in quantum batteries, emphasizing the roles of coherence, population inversion, and incoherent ergotropy. In particular, the authors showed that the ergotropy is bounded from below by the incoherent ergotropy and from above by the stored energy of the quantum battery, while also highlighting the influence of purity and energy-level populations on charging performance. Other studies have investigated quantum battery performance in superconducting qubits~\cite{intro7,intro7_a,intro7_b} and topological quantum batteries under the influence of the Zeno effect~\cite{intro8}.
	
	Furthermore, engineered reservoirs for optimal coherence-driven charging have been discussed in~\cite{intro9,intro9_a}. Moreover, the impact of non-Markovian dynamics under strong coupling on optimal charging with engineered reservoirs has been explored in~\cite{intro10}. More recently, engineered collective dissipation has been shown to generate symmetry-protected dark and metastable-like sectors in open quantum batteries, providing a mechanism for enhancing ergotropy and charging power through dissipative protection~\cite{intro10_a}. Importantly, in all the aforementioned works, the focus has been primarily on the charging process itself, without emphasizing the role of correlations or coherence in the charging dynamics. Most studies have considered energy storage assisted by external work in coherence-driven charging~\cite{intro11,intro12,intro13,intro14,intro14_a}.  
	
	In the context of reservoir engineering~\cite{INTRO_rev_1,INTRO_rev_2,INTRO_rev_3}, the reservoir acts fundamentally as a dissipative mediator, inducing energy transfer between charger and battery. It also enables analysis of wireless-like charging processes and the effects of Markovian and non-Markovian dynamics in quantum many-body systems~\cite{INTRO_rev_4,INTRO_rev_5}. Additionally, dissipative charging based on population inversion directly from the bath has been demonstrated autonomously in~\cite{INTRO_rev_6}.
	
	In our framework, we consider a theoretical model of a structured reservoir consisting of two qubits, which interact with a charger–battery architecture. We investigate the Hilbert space structure of the reservoir qubits, charger, and battery based on the global resonance condition between the subsystems. Three scenarios are analyzed: the first scenario is a direct coupling between the structured reservoir qubits and the battery; the second scenario is a common coupling between the structured reservoir qubits, charger, and battery. While the third scenario is a common coupling between the structured reservoir qubits and the charger, with a local interaction between the charger and battery. 
	
	These three scenarios are examined for two types of initial states, namely, coherent and incoherent initial states. This allows us to highlight the effect of correlations in the semiclassical case and the role of coherence in the quantum regime. Moreover, we study the impact of the structured reservoir and each scenario on the charging process by evaluating the ergotropy, stored energy and charging power of the quantum battery. Additionally, we provide a theoretical bound on the ergotropy based on the free energy stored in the coherence of the subsystems. Numerical results are presented for each scenario with the corresponding bounds on the battery’s ergotropy. This allows us to make a comparison between scenarios and identify the most efficient configuration for quantum battery charging.
	
	This distinguishes our work from previous contributions~\cite{intro6,intro7,intro8,intro9,intro10,intro11,intro12,intro13,intro14,intro14_a}. First, we investigate the impact of structured reservoirs and their qubits on the charger--battery system, allowing the quantum battery to be charged autonomously in each scenario without external work. Second, we investigate the role of coherence, population and Hilbert space structure as quantum resources on the charging process in each scenario. In particular, we focus our attention on a structured reservoir consisting of two quantum subsystems that participate coherently in the dynamics and generate correlations through resonant many-body interactions. Consequently, the charging process is not purely environment-mediated but arises from coherent energy exchange under global resonance. In addition to the theoretical model of the structured reservoir--charger--battery system, we study the theoretical bounds on ergotropy and the impact of correlations as a resource used to enhance the charging of the quantum battery. Finally, we establish upper and lower bounds on the ergotropy for each scenario.
	
	Our study interfaces with a growing literature on quantum batteries, ergotropy, and the role of coherence and population. In fact, Alimuddin $et$ $al.$ studied the effect of passive states structure on optimal charging, emphasizing that the limits of the state depend on extractable work \cite{MODEL8}. Moreover, Francica $et$ $al.$ showed that quantum coherence can affect ergotropy and derived coherence-based bounds \cite{MODEL10}. While, A. Touil $et$ $al.$ discussed the effect of classical correlations on quantum ergotropy \cite{MODEL12}. However, Biswas $et$ $al.$ further connected ergotropy extraction to free-energy bounds with potential applications in open-cycle engines \cite{MODEL13}. More recently, Castellano $et$ $al.$ generalized the notion of local ergotropy to extended settings, refining how work can be locally extracted in multipartite systems \cite{MODEL9}.

	Our paper is organized as follows. In Sec.~\ref{sec:Theoretical Model and General Framework}, we discuss the Hamiltonian model and the dynamics of our theoretical framework, including the proposed three scenarios of interaction between the structured reservoir and the charger-battery system in Sec.~\ref{sec:Quantum Battery and Structured Reservoir Scenarios}. The scenarios ($I$, $II$, $III$) are presented in Subsubsecs.~\ref{sec:Scenario I: Common interaction between $S_{12}$ and the battery}, \ref{sec:Scenario II: Common interaction between $S_{12}$ and the charger–battery system}, and \ref{sec:Scenario III: Common interaction between $S_{12}$ and the charger, and local charger–battery interaction}. In Subsec.~\ref{sec:Charging Process: Stored Energy, Ergotropy, and Contributions from Coherence and Populations}, we analyze the theoretical framework of the charging process, energy storage, and ergotropy of the quantum battery. The free energy of coherence and bounds on ergotropy of a quantum battery are discussed in Subsubsec.~\ref{sec:Free Energy of Coherence and Bounds on Quantum Battery Ergotropy}. The numerical results are presented in Sec.~\ref{sec:Results and Discussion}, where the stored energy, ergotropy, and charging power are analyzed in Subsec.~\ref{sec:Charging Process}. While, the role of coherence and population on the charging of the battery is examined in Subsec.~\ref{sec:Effects of coherence and population on the charging process}. We finish with a conclusion and perspectives in Sec.~\ref{sec:Conclusion}.
	
	\section{Theoretical Model and General Framework}\label{sec:Theoretical Model and General Framework}
	The model of interest consists of a quantum battery, denoted by $B$, connected to a quantum charger, $C$. The battery–charger system is coupled to a structured reservoir, which consists of two qubits, $S_1$ and $S_2$, each interacting with its own bosonic thermal reservoir, $R_1$ and $R_2$, respectively.  
	
	Note that many types of reservoirs can be used for $R_1$ and $R_2$. Particularly, we choose bosonic reservoirs as an illustrative example in this work. However, fermionic or squeezed reservoirs can also be considered~\cite{MODEL1a,MODEL2a}. In fact, we aim to investigate the role of correlations between the qubits in the structured reservoir, rather than focusing on the specific nature of the reservoirs.
	
\begin{figure}[h!]
		\includegraphics[scale=0.5]{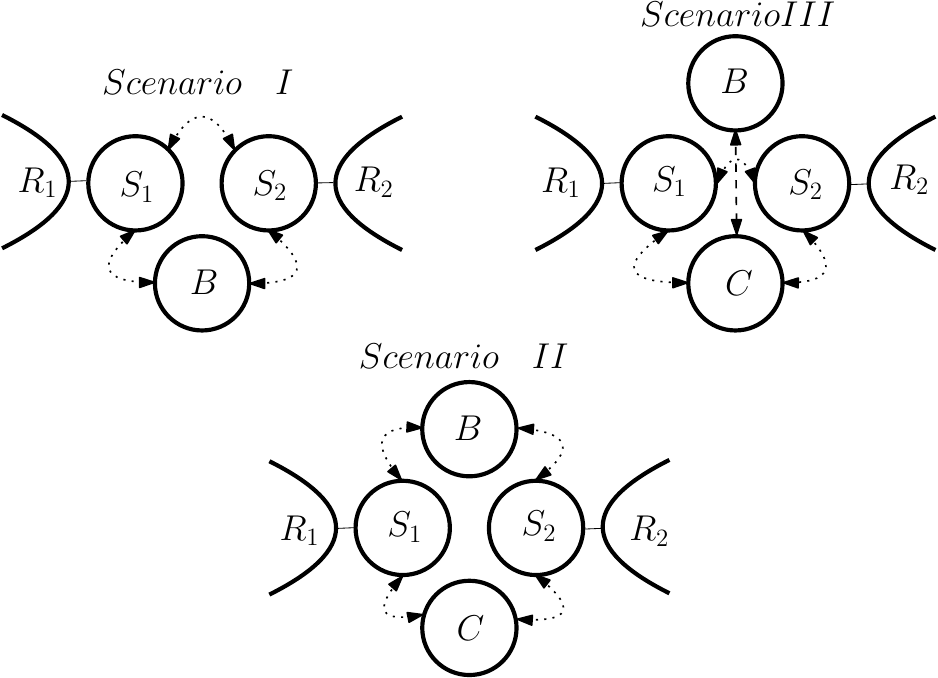}
		\caption{Schematic representation of the three interaction scenarios between the structured reservoir qubits $S_{1}$ and $S_{2}$, charger $C$, and quantum battery $B$. 
			(a) Scenario $I$: direct coupling between $S_{12}$ and $B$. 
			(b) Scenario $II$: common coupling between $S_{12}-C-B$ system. 
			(c) Scenario $III$: common coupling between $S_{12}-C$, together with a local interaction between $C-B$. 
			Each reservoir qubit $S_{m}$ is in contact with its own bosonic thermal reservoir $R_{m}$ $(m=1,2)$.}
		\label{Model}
	\end{figure}

	In our work, we will focus on the system given as $S = S_1 \otimes S_2 \otimes C \otimes B$, which is characterized using the following Hamiltonian (with $\hbar=1$ and $K_B =1$)
	\begin{align}
		\hat{H}_{S} &= \sum_{m=1}^{2} \hat{H}_{S_m} + \sum_{n=\{C,B\}} \hat{H}_n + \hat{H}_{Int_{\alpha}},\\
		\hat{H}_{S_m} &= \omega_{S_m} \hat{\sigma}_{S_m}^{+} \hat{\sigma}_{S_m}^{-}, \quad (m=\{1,2\}),\\
		\hat{H}_{n} &= \omega_{n} \hat{\sigma}_{n}^{+} \hat{\sigma}_{n}^{-}, \quad (n=\{C,B\}),
	\end{align}
	where $\hat{H}_{S_m}$ and $\hat{H}_{n}$ are the free Hamiltonians, and $\omega_{S_m}$ for ($m=\{1,2\}$) and $\omega_{n}$ for ($n=\{C,B\}$) are the energy spacings of the qubits $S_m$ and $n$, respectively. The raising and lowering operators are defined as \(\hat{\sigma}_{[.]}^+ = \ket{1}_{[.]}\bra{0}\) and \(\hat{\sigma}_{[.]}^- = \ket{0}_{[.]}\bra{1}\), acting on the computational basis states of the respective qubits. In the following subsection, we shall investigate the Hamiltonian $\hat{H}_{\mathrm{Int}_{\alpha}}$ for $\alpha = \{I, II, III\}$, which describes the interaction Hamiltonian corresponding to Scenarios~I, II, and~III. This Hamiltonian defines the interaction between the qubits of $S$. Then, we shall consider the proposed three scenarios, as illustrated in Fig.~\ref{Model}. After analyzing the dynamics of $S$, we will discuss each scenario based on the form of $\hat{H}_{\mathrm{Int}_{\alpha}}$ in the Subsec.\ref{sec:Quantum Battery and Structured Reservoir Scenarios}.\\
	
	The total density matrix, namely $\hat{\rho}_{S}(0)$ takes the following compact form:
\begin{align}
		\hat{\rho}_{S}(0) &= \hat{\rho}_{S_{12}}(0) \bigotimes_{n=\{C,B\}} \hat{\rho}_{n}(0),\\
		\hat{\rho}_{S_{12}}(0) &= \hat{\rho}_{S_1}(0) \otimes \hat{\rho}_{S_2}(0) +\hat{\rho}_{S_{12}}^{cor}(0),\nonumber
\end{align}
where $\hat{\rho}_{S_{12}}(0)$ denotes the initial state of the structured reservoir qubits $S_1$ and $S_2$. While,  $\hat{\rho}_{S_{12}}^{\mathrm{cor}}(0)$ defines the initial correlation operator between $S_1$ and $S_2$. If the two qubits are initially uncorrelated, then $\hat{\rho}_{S_{12}}^{\mathrm{cor}}(0)=\hat{0}$. Moreover, $\hat{\rho}_{n}(0)$ denotes the initial state of the qubit $n=\{C,B\}$. In our treatment, we consider that each qubit, namely $S_{1}$ and $S_{2}$, is weakly and resonantly coupled to its corresponding reservoirs, $R_{1}$ and $R_{2}$, respectively. However, since $R_{1}$ and $R_{2}$ are bosonic thermal reservoirs, the evolution of  $\hat{\rho}_{S}(t)$ is described using the following local Markovian-master equation \cite{MODEL1, MODEL2, MODEL3}
	\begin{equation}\label{MASTER_EQUATION}
		\frac{d}{d t} \hat{\rho}_S(t) = -i \left[ \sum_{m=1}^{2} \hat{H}_{S_m} + \sum_{n=\{C,B\}} \hat{H}_n, \hat{\rho}_S(t) \right] + \mathcal{L}_{S_{12}}[\hat{\rho}_S(t)],
	\end{equation}
	where $[\,\cdot\,,\,\cdot\,]$ denotes the usual commutator. The first term in the right-hand of Eq.~\ref{MASTER_EQUATION} gives rise to the free evolution of the system under its free Hamiltonians. Indeed, it corresponds to the reversible (unitary) evolution of the system $S$. The second term represents the interaction between the subsystems of $S$ and the dissipative interaction of qubits $S_m$ with their respective reservoirs $R_m$. This term accounts for the irreversible (non-unitary) evolution of the total system $S$, which is given as below: 
	\begin{align}
		\mathcal{L}_{S_{12}}(t)[\hat{\rho}(t)]=&-i\left[H_{Int_{\alpha}}, \hat{\rho}_S(t)\right]+\sum_{m=1}^{2}\mathcal{D}^{[T_{R_m}]}[\hat{\rho}_S(t)],\label{LL}\\
		\mathcal{D}^{[T_{R_m}]}[\hat{\rho}_S(t)]=&\gamma_{m}(\bar{n}_{m}(T_{R_m},\omega_{S_m})+1)\mathcal{D}^{[\hat{\sigma}_{S_{m}}^{-}]}[\hat{\rho_S}(t)]+\nonumber\\
		&\gamma_{m}\bar{n}_{m}(T_{R_m},\omega_{S_m})\mathcal{D}^{[\hat{\sigma}_{S_{m}}^{+}]}[\hat{\rho}_S(t)]\nonumber,\\
		\mathcal{D}^{[\hat{\sigma}_{S_{m}}^{\pm}]}[\hat{\rho}_S(t)]&=\hat{\sigma}_{S_{m}}^{\pm}\hat{\rho_S}(t)\hat{\sigma}_{S_{m}}^{\mp}-\frac{1}{2}\{\hat{\sigma}_{S_{m}}^{\mp}\hat{\sigma}_{S_{m}}^{\pm},\hat{\rho_S}(t)\},\nonumber
	\end{align}
	
	where, $T_{R_{m}}$ for $m=\{1,2\}$ denote the temperatures of the reservoirs. While, $\gamma_{m}$ are the decay rates of the reservoirs $R_{m}$, which set the timescale of dissipation. Under the resonance condition, namely rotating-wave approximation \cite{MODEL3a} and in the weak-coupling limit approximation, this description is valid and it provides that $\gamma_{m} \ll \omega_{S_{m}}$. However, the quantity $\bar{n}_{m}(T_{R_m},\omega_{S_m})$ represents the average number of quanta (bosons) in the reservoir $R_m$. In fact, it is given as:
	\begin{align*}
		\bar{n}_{m}(T_{R_m},\omega_{S_m}) = \frac{1}{\exp\left(\frac{\omega_{S_m}}{T_{R_m}}\right) - 1}.
	\end{align*}
	For sake of simplicity, we consider that $T_{R_1} = T_{R_2} = T$ throughout the paper. In the following, we investigate the interaction Hamiltonian $\hat{H}_{\mathrm{Int}_{\alpha}}$  to define each scenario of the interaction between structured reservoir qubits, charger and battery coupling.
	
	\subsection{Quantum Battery and Structured Reservoir Scenarios}
	\label{sec:Quantum Battery and Structured Reservoir Scenarios}
	Structured reservoirs play an important role in the study of backflow effects and energy transfer in the battery–charger system and auxiliary (ancilla) qubits of the structured reservoir \cite{MODEL4, MODEL5,Oularabi2025,Aiache2025}. In our case, the structured reservoir consists of two bosonic reservoirs and two qubits, namely $S_1$ and $S_2$, which interact with the external charger–battery system. Here, the exchange of correlations between the subsystems and the transfer of coherence play an important role in the control and enhancement of the charging process. Then, we will investigate the Hilbert space of the system $S=S_1\otimes S_2 \otimes C\otimes B$ qubits to realize the proposed three scenarios of the interaction Hamiltonian $\hat{H}_{Int}$ between the structured reservoir qubits $S_{12}=S_1 \otimes S_2$ and the charger-battery subsystem.
	\subsubsection{Scenario I: Common interaction between $S_{12}$ and the battery}
	\label{sec:Scenario I: Common interaction between $S_{12}$ and the battery}
	For this scenario, we consider that the quantum battery $B$ is in direct contact with the qubits $S_{12}$, as illustrated in Fig.~\ref{Model} (Scenario~I). In fact, we aim to investigate a common interaction between $S_1$, $S_2$, and $B$. In this scenario, such an interaction allows the quantum battery to be charged autonomously from $S_{12}$ without any external work and charger. In this scenario, the main goal is to induce transitions from the state $\ket{0_{S_1} 1_{S_2} 0_B}$ to that $\ket{1_{S_1} 0_{S_2} 1_B}$, which drives effectively the battery from its passive state $\ket{0_B}$ to the active one $\ket{1_B}$ through a population inversion, under the resonance condition between $S_{12}$ and the battery. Indeed, it satisfies the following condition:
	
	\begin{align}\label{resI}
		\omega_B = \omega_{S_2} - \omega_{S_1},\quad \omega_{S_2}>\omega_{S_1},
	\end{align}
	
	which physically corresponds to the exchange of excitation between the subsystem qubits $S = S_1 \otimes S_2 \otimes B$. Therefore, the interaction Hamiltonian for this scenario, namely $\hat{H}_{\text{Int}_I}$ takes the following form:
	
	\begin{align}\label{INT1}
		\hat{H}_{\text{Int}_I} = g_{1} \left[ \ket{0_{S_1}1_{S_2}0_B}\bra{1_{S_1}0_{S_2}1_B} + \text{H.c.} \right],
	\end{align}
	
	where $g_1$ is the coupling between $B$ and $S_{12}$. In this configuration, $S_{12}$ acts as a filter for the decoherence effects of the reservoirs $R_1$ and $R_2$ on the quantum battery. Also, it is used in the context of autonomous quantum thermal machines in \cite{MODEL6,MODEL7}. Note that the coupling coefficient satisfies $g_1 \ll \omega_B$ to ensure the validity of the weak-coupling limit in the master equation already given in Eq.~\ref{MASTER_EQUATION}. 
	
	An experimental realization of this scenario has been reported by McKay $et$ $al.$ \cite{MODEL4a}. Indeed, they provided a practical method for controlling the parameters with high precision in superconducting qubits. These techniques can be applied to the qubits $S_{1}$, $S_{2}$ and $B$ using, for example, three coupled transmon qubits via a tunable bus resonator in a superconducting cavity \cite{MODEL5a}.
	\subsubsection{Scenario II: Common interaction between $S_{12}$ and the charger–battery system}
	\label{sec:Scenario II: Common interaction between $S_{12}$ and the charger–battery system}
	In this scenario, we analyze the case of a common interaction between $S_{12}$, $C$ and $B$, as illustrated in Fig.~\ref{Model} (Scenario~II). This common interaction between the structured reservoir qubits, charger and battery, which corresponds to a collective excitation exchange represented as a four-body correlated transition, enables simultaneous energy transfer along the chain $S_{12} \!-\! C \!-\! B$. In fact, we realize a common interaction between the total system qubits to drive the transition from $\ket{0_{S_1}1_{S_2}1_C0_B}$ to $\ket{1_{S_1}0_{S_2}0_C1_B}$ under the following resonance condition: 
\begin{align}\label{RES_II}
		\omega_{S_2} - \omega_{S_1} = \omega_{B} - \omega_{C},
\end{align}
	where $\omega_{S_2} > \omega_{S_1}$ and $\omega_{C} < \omega_{B}$, which physically corresponds to the exchange of excitation between the total system qubits $S$. In this scenario, $S_{12}$ filters the decoherence effects of $R_1$ and $R_2$ on the charger-battery system. Then, the corresponding interaction Hamiltonian, namely $\hat{H}_{\text{Int}_{II}}$ is : 
	\begin{align}\label{INT2}
		\hat{H}_{\text{Int}_{II}} = g_{2} \left[ \ket{0_{S_1}1_{S_2}1_C0_B}\bra{1_{S_1}0_{S_2}0_C1_B} + \text{H.c.} \right],
	\end{align}
	
	where, $g_2$ denotes the coupling between $S_{12}$ and the charger–battery system. In the weak-coupling limit, this interaction satisfies the condition $g_2 \ll (\omega_{S_2} - \omega_{S_1})$. This scenario can also be applied in the context of autonomous quantum thermal machines for entanglement generation \cite{MODEL6a}. Moreover, it can be used in the case of quantum many-body systems for engineering four-qubit coherent interactions \cite{MODEL7a}. For the experimental implementation, the qubits $S_1$, $S_2$, charger and battery can be connected via a tunable coupler or a multi-mode resonator \cite{MODEL8a}, which allows the control of multi-qubit interactions through tunable couplings \cite{MODEL9a}.
	
	\subsubsection{Scenario III: Common interaction between $S_{12}$ and the charger, and local charger–battery interaction}
	\label{sec:Scenario III: Common interaction between $S_{12}$ and the charger, and local charger–battery interaction}
	Now, let consider a common interaction between $S_{12}$ and charger to drive the transition from the state $\ket{0_{S_1}1_{S_2}0_C}\bra{0_{S_1}1_{S_2}0_C}\otimes \mathcal{I}_{B}$ to that $\ket{1_{S_1}0_{S_2}1_C}\bra{1_{S_1}0_{S_2}1_C}\otimes \mathcal{I}_{B}$. Moreover, the interaction between charger and quantum battery biases the transition from  $\mathcal{I}_{S_{12}}\otimes\ket{1_C0_B}\bra{1_C0_B}$ to  $\mathcal{I}_{S_{12}}\otimes\ket{0_C1_B}\bra{0_C1_B}$ under the following resonance condition: 
	\begin{align}\label{resIII}
		\omega_B = \omega_C = \omega_{S_2} - \omega_{S_1}, \quad \omega_{S_2} > \omega_{S_1},
	\end{align}
	
	Physically, the condition in Eq.(\ref{resIII}) corresponds to the sequential transfer of energy from $S_{12}$ to the charger, and from charger to quantum battery, mediated by the interaction between $S_{12}$ and $C$. This scenario gives rise to a collective interaction between $S_{12} = S_1 \otimes S_2$ and $C$, which is similar to Scenario~I, and the local coupling between charger and battery. More precisely, in this scenario, we highlight the effect of including the charger system as a part of the structured reservoir on the charging process of quantum battery $B$. Hence, the corresponding interaction Hamiltonian $\hat{H}_{\text{Int}_{III}}$ is given follows:  
	\begin{align}\label{INT3}
		\hat{H}_{\text{Int}_{III}} = & g_{3}\left[ \ket{0_{S_1}1_{S_2}0_C}\bra{1_{S_1}0_{S_2}1_C} + \text{H.c.} \right]\otimes \mathcal{I}_{B} \nonumber\\
		& + k\, \mathcal{I}_{S_{12}} \otimes \left[\ket{1_C0_B}\bra{0_C1_B} + \text{H.c.} \right],
	\end{align}
	where, $g_3$ and $k$ are the respective coupling strengths for the $S_{12}$–charger and charger–battery interactions. In this case, $S_{12} \otimes C$ filters the decoherence from $R_1$ and $R_2$ before it reaches the battery. Moreover, we should note that these couplings satisfy the conditions $g_3 \ll \omega_{S_2} - \omega_{S_1}$ and $k \ll \omega_{S_2} - \omega_{S_1}$, ensuring the validity of the weak-coupling approximation in the local master equation.
	For the experimental implementation, the collective coupling between $S_{12}$ and the charger through a shared resonator (similar to Scenario I), with $g_3$ is controlled via flux bias or frequency tuning. While the charger-battery coupling is locally used as a direct capacitive link or fixed-frequency transmons connected via a tunable coupler \cite{MODEL4a, MODEL5a}.
	
	Note that the resonance conditions between the subsystems, given in Eqs.~\ref{resI}, \ref{RES_II}, and \ref{resIII}, correspond to energy-matching relations between the initial and final many-body states connected by the interaction Hamiltonian $\hat{H}{\mathrm{Int}\alpha}$. These conditions ensure energy conservation during the exchange of excitations among the subsystems in all scenarios ($I$, $II$, $III$), and thus describe excitation transfer within the total system of qubits $S$.
			This dynamics can be effectively described within a two-level subspace spanned by the resonantly connected collective states, leading to identical Rabi frequencies up to a rescaling of the effective coupling.
	
	In the following sections of the paper, we analyze the impact of each scenario on the charging process, as well as the influence of coherence and memory effects on the maximum work extracted from the battery over time.
	\subsection{Charging Process: Stored Energy, Ergotropy, and Contributions from Coherence and Populations}
	\label{sec:Charging Process: Stored Energy, Ergotropy, and Contributions from Coherence and Populations}
	The interaction Hamiltonian, namely $\hat{H}_{\mathrm{Int}_\alpha}$ plays a central role in the dynamics of the total system $S$. Specifically, according to $\mathcal{L}_{S_{12}}(t)[\hat{\rho}(t)]$ in Eq.~\ref{LL}, the commutator $-i\left[\hat{H}_{\mathrm{Int}_\alpha}, \hat{\rho}_S(t)\right]$ governs both the generation of correlations and the redistribution of coherence among the subsystems $S_1$, $S_2$, $C$ and $B$ over time. However, note that the structured reservoir qubits, $S_1$ and $S_2$ do not eliminate decoherence, but they modify the effective dissipative channels. In particular, the interaction Hamiltonian facilitates the generation of correlations between the subsystems.  
	
	Based on this framework, we analyze the charging process of the quantum battery in each scenario ($I$, $II$, and $III$) by evaluating the stored energy, ergotropy, and free energy of coherence. Furthermore, we examine the bounds on quantum ergotropy and the influence of global and local coherence, as well as correlations, during the charging process.
	\subsubsection{ Energy Storage and Ergotropy of the Quantum Battery}
	\label{sec:Energy Storage and Ergotropy of the Quantum Battery}
	
	The energy stored in the quantum battery $B$ over time,
	namely  $E_{B}$ is a key quantity used to highlight the energy transmitted to the quantum battery \cite{MODEL15}. Mathematically, it is defined as follows:
	\begin{align}\label{ENERGY}
		E_{B}=Tr\{\hat{H}_{B}\hat{\rho}_B \}. 
	\end{align}

	Note that the above energy in Eq.(\ref{ENERGY}) can increase over time, which does not necessarily mean that the battery is charged. Physically, it represents only the stored energy. Indeed, the quantum battery can be charged only if its
	state  is active (non-passive)~\cite{MODEL8}, meaning that work can be extracted
	from it. To analyze the maximum extractable work from the quantum battery, we use the concept of quantum ergotropy~\cite{MODEL9}, that is, $\mathcal{E}_B$, which defines the difference between the stored energy and the energy of the quantum battery in its passive state, namely $E_{B_\mathrm{Pss}}$. Indeed, the ergotropy is defined as follows~\cite{MODEL9} 
	\begin{align}\label{ERGOTROPY_eqt}
		\mathcal{E}_B = E_{B} - E_{B_\mathrm{Pss}}, 
	\end{align}
	where $E_{B_\mathrm{Pss}}$ is the energy of the quantum battery in its passive state, which takes the following form: 
	\begin{equation}\label{er}
		E_{B_\mathrm{Pss}} = \mathrm{Tr}\{\hat{H}_{B}\,\hat{\rho}_{B_\mathrm{Pss}}\},
	\end{equation}
	where $\hat{\rho}_{B_\mathrm{Pss}}$ denotes the passive state of quantum battery~\cite{MODEL9}. Mathematically, the energy of the ground state of the quantum battery is taken as zero, while the energy of the excited state is $\omega_B$. Then, the spectral decomposition of the quantum battery state is given as:
	\begin{align}
		\hat{\rho}_B = \sum_{i=0}^{1} \psi_i \ket{\psi_i}\bra{\psi_i},
	\end{align}
	where $\psi_1(t) \leq \psi_0(t)$ and $\ket{\psi_i}$ are the eigenvalues and eigenvectors of $\hat{\rho}_B$. Then, the corresponding passive state already given in Eq.\ref{er} is defined as:
	\begin{align}
		\hat{\rho}_{B_\mathrm{Pss}} = \sum_{i=0}^{1} \psi_i \ket{i}\bra{i}.
	\end{align}
	The ergotropy of quantum battery in Eq.~\ref{ERGOTROPY_eqt} can consists also of two contributions~\cite{MODEL10,MODEL11,MODEL12,MODEL13}. The first contribution is the so-called population ergotropy $\mathcal{E}_B^{P}$, which represents the maximal extractable work due to the population distribution of quantum battery $B$. While the second contribution gives rise to the coherence ergotropy, namely $\mathcal{E}_B^{C}$, which reflects the maximal extractable work due to the quantum coherence of the quantum battery. Mathematically, the ergotropy in Eq.~\ref{ERGOTROPY_eqt} can be expressed as bellows \cite{MODEL10}: 
	\begin{align}\label{ERGOTROPY_eqt1}
		\mathcal{E}_B = \mathcal{E}_B^{P} + \mathcal{E}_B^{C},
	\end{align}
	where,	
	\begin{align}
		\mathcal{E}_B^{C} &= \mathcal{E}_B - \mathcal{E}_B^{P},\label{ERGO_COH}\\
		\mathcal{E}_B^{P} &= E_B - \bar{E}_{B_\mathrm{Pss}},\label{ERGO_POP} 
	\end{align}
	where $\bar{E}_{B_\mathrm{Pss}} = \mathrm{Tr}\{\hat{H}_B \bar{\rho}_{B_\mathrm{Pss}}\}$ defines the energy of the passive state $\bar{\rho}_{B_\mathrm{Pss}}$ of the fully dephased state of the quantum battery~\cite{MODEL14}, denoted $\bar{\rho}_B$. Mathematically, it is given as: 
	\begin{align}
		\bar{\rho}_B = \sum_{i=0}^{1} \braket{i|\hat{\rho}_B|i} \ket{i}\bra{i}.
	\end{align}
	For each subsystem, $\bar{\rho}_{x}$ for $x=\{S_1, S_2, C\}$ denotes the corresponding incoherent state. In the following part, we shall describe the bounds on the quantum ergotropy and the impact of global and local coherence on the charging process of quantum battery.
	\subsubsection{Free Energy of Coherence and Bounds on Quantum Battery Ergotropy}
	\label{sec:Free Energy of Coherence and Bounds on Quantum Battery Ergotropy}
	As demonstrated in Eq.~\ref{ERGOTROPY_eqt}, the ergotropy can originate either from the population or coherence of the quantum battery. In the literature, the energy stored in coherence is quantified using the free energy of coherence, namely $W(\hat{\rho})$~\cite{MODEL12, MODEL13, MODEL14}. Indeed, for any state $\hat{\rho}$, the quantity $W(\hat{\rho})$ is mathematically defined as follows: 
	\begin{align}\label{FREE_ENERGY_OF_COHERENCE}
		W(\hat{\rho}) = F(\hat{\rho}) - F(\bar{\rho}) = K_B T \mathcal{C}(\hat{\rho}),
	\end{align}
	where $F(\hat{\rho})$ and $F(\bar{\rho})$ represent the free energies of the state $\hat{\rho}$ and its fully dephased state $\bar{\rho}$, respectively. While, $\mathcal{C}(\hat{\rho})$ is the relative entropy of coherence~\cite{MODEL16}, which quantifies the coherence of $\hat{\rho}$. Mathematically, the free energy of $\hat{\rho}$ is defined as~\cite{MODEL15} 
	\begin{align}
		F(\hat{\rho}) = E - K_{B} T S(\hat{\rho}),
	\end{align} 
	where $E = \mathrm{Tr}\{\hat{\rho}\hat{H}\}$ is the energy of $\hat{\rho}$ with Hamiltonian $\hat{H}$, and $S(\hat{\rho}) = -\mathrm{Tr}\{\hat{\rho}\log_2 \hat{\rho}\}$ is the von Neumann entropy. Here, $T$ denotes the reservoir temperature and $K_{B}$ is the Boltzmann constant. However, the relative entropy of coherence in Eq.\ref{FREE_ENERGY_OF_COHERENCE} is defined as~\cite{MODEL16}  
	\begin{align}\label{REC}
		\mathcal{C}(\hat{\rho}) = S(\bar{\rho}) - S(\hat{\rho}).
	\end{align}
	
	In our scenario, the free energy of coherence of battery, namely $W(\hat{\rho}_B)$, bounds the coherence ergotropy $\mathcal{E}_B^C$ as~\cite{MODEL10, MODEL12} (see Appendix~\ref{Theoretical}) 
	
	\begin{align}\label{COHERENCE BOUND}
		0 \leq \mathcal{E}_B^{C} \leq W(\hat{\rho}_B) := K_B T\, \mathcal{C}(\hat{\rho}_B),
	\end{align}
	
This result implies that the maximum extractable work from coherence is bounded by the free energy of coherence. It represents the maximal work that can be stored in coherence, indicating that the full coherent energy cannot be extracted due to decoherence effects induced by the reservoir. Note that Eq.~\eqref{COHERENCE BOUND} is not universal since it depends on the inequality given in Eq.~\eqref{CONDITION_VALIDITY}. However, note that more universal bounds include the contribution based on the free energy stored with respect to the equilibrium state~\cite{FINAL1,FINAL2,FINAL3,FINALE}. Our bound highlights the contribution of global and local coherences, as well as the exchange of correlations between the subsystems with respect to the production of ergotropy. Besides, note that the bound in Eq.~\ref{COHERENCE BOUND} characterizes directly the maximum extractable work associated to coherence. However, in our theoretical model, the energy transfer is mediated by the structured reservoir qubits $S_1$ and $S_2$, as well as the charger $C$. Therefore, this bound does not fully capture the global coherence, local contributions of the subsystems, or correlations generated by the interaction Hamiltonian between them. Now, by using Eqs.~\ref{ERGOTROPY_eqt1} and \ref{COHERENCE BOUND}, one can show that the total ergotropy of the quantum
	battery is also bound as:
	
	\begin{align}\label{Ergo BOUND}
		\mathcal{E}_B^{P} \leq \mathcal{E}_B \leq K_B T \mathcal{C}(\hat{\rho}_B) + \mathcal{E}_B^{P}.
	\end{align}
	
	Thus, there are two bounds on the quantum battery ergotropy, namely an upper bound: $\mathcal{E}_B \leq K_B T\, \mathcal{C}(\hat{\rho}_B) + \mathcal{E}_B^{P}$ and a lower bound given as $\mathcal{E}_B \geq \mathcal{E}_B^{P}$. This implies that the minimal value of the ergotropy is $\min \mathcal{E}_B = \mathcal{E}_B^{P}$, while the maximal value is $\max \mathcal{E}_B = K_B T\, \mathcal{C}(\hat{\rho}_B) + \mathcal{E}_B^{P}$.

	Note that if the ergotropy of population is zero, the ergotropy is bounded between $\min \mathcal{E}_B = 0$ and $\max \mathcal{E}_B = K_B T\, \mathcal{C}(\hat{\rho}_B)$. Conversely, for an incoherent state where $K_B T\, \mathcal{C}(\hat{\rho}_B) = 0$, the ergotropy arises solely from the population, such that $\mathcal{E}_B = \mathcal{E}_B^{P}$. These results are also provided in~\cite{MODEL12,MODEL13,MODEL14}, which give the upper and lower bounds of the ergotropy.\\

	Moreover, the global and local coherence of the subsystems, i.e., structured reservoir qubits $S_{12} = S_1 \otimes S_2$, charger, and quantum battery, which represent continuous resources for the charging process (Eq.~\ref{Ergo BOUND}). The difference between global and local coherence, namely $\Delta C(\hat{\rho}_S)$ is defined as follows: 
	\begin{align}
		\Delta C(\hat{\rho}_S) &= C(\hat{\rho}_S) - \sum_{m=1}^{2} \mathcal{C}(\hat{\rho}_{S_m}) - \sum_{n=\{C,B\}} \mathcal{C}(\hat{\rho}_n) \nonumber\\
		&= I_S - \bar{I}_S, \label{DELTAC}
	\end{align}
	where $I_S$ is the conventional mutual information~\cite{MODEL17_correct} of the total system $S$, describing the correlations (quantum and classical) between the subsystems $S_1$, $S_2$, $C$, and $B$. It is given as: 
	\begin{align}
		I_S = \sum_{m=1}^{2} S(\hat{\rho}_{S_m}) + \sum_{n=\{C,B\}} S(\hat{\rho}_n) - S(\hat{\rho}_S).
	\end{align}
	While, $\bar{I}_S$ in Eq.\ref{DELTAC} gives rise to the conventional mutual information of the fully dephased state of the total system $S$, describing correlations due to the populations in the subsystems of $S$. This approach is also discussed in~\cite{MODEL12, MODEL13}, where mutual information is used for quantum systems with only two subsystems.\\	
	In our case, a general bound on the ergotropy of a quantum battery can be obtained based on the total relative entropy of $S$. Hence, by using Eq.~\ref{REC}, the free energy of coherence of the total system $S$ is given as bellows: 
	\begin{align}
		W(\hat{\rho}_S) = K_B T \, \mathcal{C}(\hat{\rho}_S).
	\end{align}
	Thus, the free energy of coherence of the total system $S$ and that of the quantum battery $B$ can be expressed as:
	\begin{align}\label{FREE_ENERGY_OF_COHERENCE_B}
		W(\hat{\rho}_S) = K_B T \left( \sum_{m=1}^{2} \mathcal{C}(\hat{\rho}_{S_m}) + \sum_{n=\{C,B\}} \mathcal{C}(\hat{\rho}_n) + I_S - \bar{I}_S \right),\nonumber\\
		W(\hat{\rho}_B)= W(\hat{\rho}_S) -  K_B T \left( \sum_{m=1}^{2} \mathcal{C}(\hat{\rho}_{S_m}) + \mathcal{C}(\hat{\rho}_C) + I_S - \bar{I}_S \right).
	\end{align}
	Note that the whole free energy of coherence is given as the sum of the free energy of coherence of each subsystem, namely $W(\hat{\rho}_S)- K_B T \left( \sum_{m=1}^{2} \mathcal{C}(\hat{\rho}_{S_m}) + \mathcal{C}(\hat{\rho}_C)\right)$, plus the difference between total and global relative entropy of coherence of the total system, that is, $I_S - \bar{I}_S$. In our case, we use the conventional mutual information because we study four interacting subsystems. Therefore, $\Delta C(\hat{\rho}_S)$ demonstrates that the ergotropy in our battery is affected by the total correlation exchange among the subsystems of $S$.

	Moreover, the free energy of coherence of the quantum battery $W(\hat{\rho}_B)$ in Eq.~\ref{FREE_ENERGY_OF_COHERENCE_B} highlights the effect of coherence transfer between the subsystems $S_{12}=S_1 \otimes S_2$, charger and battery. Also, it captures the difference between global and local coherence as a quantum resource for coherence-based energy storage. Note that the free energy of coherence consists of the difference between two contributions, namely the energy of coherence and the energy of correlations, which are expressed respectively as follows: 
	\begin{eqnarray}
		W(\hat{\rho}_B) &= &\mathcal{W}_{coh} - \mathcal{W}_{cor}, \nonumber\\
		\mathcal{W}_{coh} &=& K_B T \left( \mathcal{C}(\hat{\rho}_S) - \sum_{m=1}^{2} \mathcal{C}(\hat{\rho}_{S_m}) - \mathcal{C}(\hat{\rho}_C) \right), \nonumber\\
		\mathcal{W}_{cor} &=& K_B T \left( I_S - \bar{I}_S \right),		
	\end{eqnarray}
	where the contribution $\mathcal{W}_{coh}$ arises from the exchange of coherence between the subsystems, the structured reservoir qubits $S_{12}=S_1 \otimes S_2$, charger and quantum battery. However, the correlation contribution by means of $\mathcal{W}_{Cor}$ quantifies the amount of information and memory effects in the system. Here, coherence represents the energy stored in coherent superpositions, while correlations represent the consumed energy required to preserve information shared between the subsystems, allowing the control over decoherence effects from the reservoirs $R_1-S_1$ and $R_2-S_2$ in the external charger–battery system.\\
	Importantly, in our case, note that the expression in Eq.~\ref{FREE_ENERGY_OF_COHERENCE_B} can be generalized to $N$ reservoirs with $N$ structured qubits, such that $S_{N} = \bigotimes_{m=1}^{N} S_{m}$ as below: 
	\begin{equation}
		W(\hat{\rho}_B) = W(\hat{\rho}_S) 
		- K_B T \left( 
		\sum_{m=1}^{N} \mathcal{C}(\hat{\rho}_{S_m}) 
		+ \mathcal{C}(\hat{\rho}_C) 
		+ I_S 
		- \bar{I}_S 
		\right),
	\end{equation}
	where the total state is $\hat{\rho}_S = \bigotimes_{m=1}^{N} \hat{\rho}_{S_m} \otimes \hat{\rho}_C \otimes \hat{\rho}_B$. We used the example with $N = 2$ only to simplify the dynamics.\\
	Note that energy is stored in the coherence of the quantum battery if and only if the coherence energy exceeds the correlation energy (see numerical analysis in Appendix~\ref{Numerical33}). Accordingly, the bound on coherence-based energy storage is given as follows: 
		\begin{equation}\label{COH_COR_Bound}
			\mathcal{W}_{\mathrm{coh}} > \mathcal{W}_{\mathrm{cor}}.
		\end{equation}
		Physically, this bound indicates that correlations play two important roles. For $\mathcal{W}_{\mathrm{coh}} = \mathcal{W}_{\mathrm{cor}}$, all the coherence energy is consumed by correlations, which act then as an information reservoir ensuring memory effects (backflow of information between the subsystems). For $\mathcal{W}_{\mathrm{coh}} > \mathcal{W}_{\mathrm{cor}}$, correlations act as a mediator of coherence transfer between subsystems and enhance work extraction through the contribution of coherence.
	
	The corresponding bound on the ergotropy of the quantum battery $\mathcal{E}_B$ is therefore expressed as (see numerical analysis in Appendix~\ref{Numerical34}):
	\begin{align}\label{RGO_COH_COR_Bounds}
		\mathcal{E}_B^{P} \leq \mathcal{E}_B \leq \mathcal{E}_B^{P} + \mathcal{W}_{coh} - \mathcal{W}_{cor}.
	\end{align}
	This bound shows that the total ergotropy of the quantum battery is influenced by both coherence transfer and population excitation exchange. The energy associated with memory effects is dissipated through the exchange of correlations between the external subsystems. 
	Compared to the bound in Eq.~\ref{COHERENCE BOUND}, the bound in Eq.~\ref{RGO_COH_COR_Bounds} provides more detailed information about the maximal and minimal values of the total ergotropy of the quantum battery. Further, it indicates that both coherence and correlations enhance the maximal extractable work of the battery. Moreover, the correlations generated through the interaction Hamiltonian between the subsystems play an important role as a resource for coherence redistribution and decoherence filtering within the system.
	\section{Results and Discussion}
	\label{sec:Results and Discussion}
	In this section, we will investigate numerically the impact of populations, coherence, and correlations on the charging process in each scenario (I, II, and III), as well as the validity of ergotropy bounds provided in Eqs.~(\ref{COH_COR_Bound}) and (\ref{RGO_COH_COR_Bounds}). However, we study two examples of initial states for each model and make a comparison between them.
	
	In the first example given for each scenario as ($I_a$, $II_a$, $III_a$), we consider the initial state of the total system $S$ as 
	\[
	\hat{\rho}_{S_a}(0)=\ket{0_{S_1}1_{S_2}0_B}\bra{0_{S_1}1_{S_2}0_B} \quad \text{for } (I_a),
	\]
	and 
	\[
	\hat{\rho}_{S_a}(0)=\ket{0_{S_1}1_{S_2}1_C0_B}\bra{0_{S_1}1_{S_2}1_C0_B} \quad \text{for  } (II_a \text{ and } III_a).
	\]
	
	However, in the second example for each scenario ($I_b$, $II_b$, $III_b$), we consider the following state for $S_{12}$ system: 
	\[
	\ket{\psi_{S_{12}}(0)}=\frac{1}{\sqrt{2}}\Big[\ket{0_{S_1}1_{S_2}}+\ket{1_{S_1}0_{S_2}}\Big]. 
	\]
	While, for the charger system, we assume that the state is given as below: 
	\[
	\ket{\psi_C(0)}=\frac{1}{\sqrt{2}}\Big[\ket{0_C}+\ket{1_C}\Big],
	\]
	and for a quantum battery, we have: 
	\[
	\ket{\psi_B(0)}=\ket{0_B},
	\]
	where the total initial state for scenario  $I_a$ is given as: 
	\[
	\hat{\rho}_{S_b}(0)=\ket{\psi_{S_{12}}(0)\psi_B(0)}\bra{\psi_{S_{12}}(0)\psi_B(0)},
	\]
	and the total initial state for  for scenarios  ($II_a$ and $III_a$) is proposed as : 
	\[
	\hat{\rho}_{S_b}(0)=\ket{\psi_{S_{12}}(0)\psi_C(0)\psi_B(0)}\bra{\psi_{S_{12}}(0)\psi_C(0)\psi_B(0)},
	\].
	
	The example of the initial state $S_a$ highlights the impact of population on the charging process, since the initial state is incoherent (a semiclassical state). Indeed, it is used to highlight the effects of population inversion \cite{Model25,Model26} on the charging process, as well as the role of classical correlations, in the absence of any coherence exchange between the subsystems. While the example $S_b$ highlights the effect of coherence on the charging process, since the structured reservoir qubits are entangled \cite{Model27}. In fact, we analyze the effect of quantum correlation exchange on the ergotropy bound, as well as the ability of coherence transfer to enhance the charging process. For each example, note that the battery is initially prepared in a passive state with zero initial energy.
	
	In our numerical simulation, we set the energy spacing of qubit $S_2$ to $\omega_{S_2}=10$, and for the other parameter, we use $\omega_{S_1}=0.5\omega_{S_2}$. The reservoir temperatures are set as $T_{R_1}=T_{R_2}=T=\omega_{S_1}$, and the decay rate is chosen as $\gamma_{m}=0.01\omega_{S_m}$ for $m=\{1,2\}$. Moreover, we focus on analyzing the coupling between $S_{12}$ and the charger-battery system. Throughout the rest of this paper, the coupling $g$ takes the values $0.03$, $0.05$, $0.07$, and $0.09$ in units of $\omega_{S_2}$, which are represented in the plots by red dotted, black dashed, blue dotted-dashed, and magenta solid lines. For scenario III, the coupling between charger and quantum battery is chosen as $k=0.03\,\omega_{C}$.
	
	For these parameters, the condition $g/\omega_{S_2}\le 0.1$ is satisfied, which corresponds to the weak coupling limit. Experimentally, this lies within the range of superconducting qubits, where $\omega_{S_2}$ is given in the order of GHz. While, the couplings $g$ and $k$ are in the order of MHz, and the time scale in the range of $\mu s–ms$. This shows that our theoretical model can be experimentally realized using superconducting qubits~\cite{MODEL4a,MODEL5a,MODEL6a,MODEL7a,MODEL8a,MODEL9a,MODEL17,MODEL18,MODEL19,MODEL20}.
	\subsection{Charging Process}
	\label{sec:Charging Process}
	In this section, we manipulate numerically the amounts of ergotropy and energy of the quantum battery already defined in Eqs.~\ref{ERGOTROPY_eqt} and \ref{ENERGY}, respectively. In addition, we define the charging power of the battery over
	time, namely {$\mathcal{P}_B(t)$ as \cite{MMODL21}
	\begin{align}
		\mathcal{P}_B(t) = \frac{\mathcal{E}_B(t)}{t}.
	\end{align} 
	The above expression of charging power describes how fast the quantum battery becomes charged over time. Note that we use the normalized quantities$\frac{\mathcal{E}_B(t)}{\omega_{B}}$, $\frac{E_B(t)}{\omega_{B}}$, and $\frac{\mathcal{P}_B(t)}{\omega_{B}}$ to compare correctly between the three scenarios in the two examples of the initial states.
	
	\begin{figure*}[htp!]
		
		\subfloat[ \label{CHARGING_EXEMPLE_1_010}]{%
			\includegraphics[width=0.69\columnwidth]{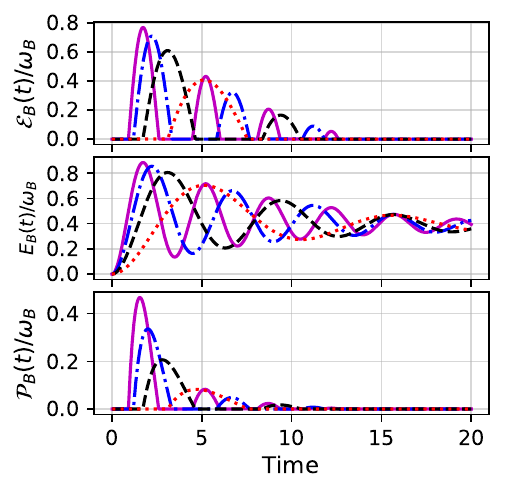}%
		}\hfill
		\subfloat[ \label{CHARGING_EXEMPLE_2_0110}]{%
			\includegraphics[width=0.69\columnwidth]{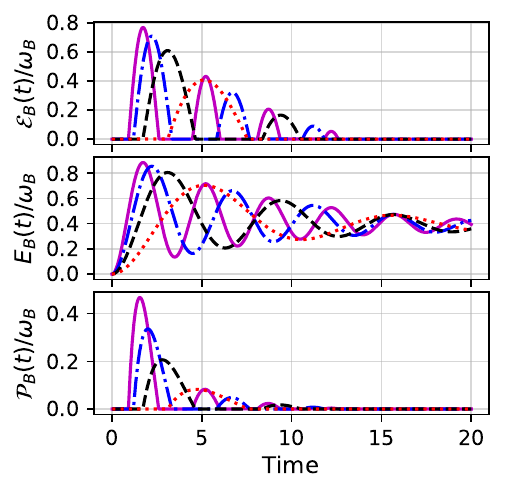}%
		}\hfill
		\subfloat[\label{CHARGING_EXEMPLE_3_0110}]{%
			\includegraphics[width=0.69\columnwidth]{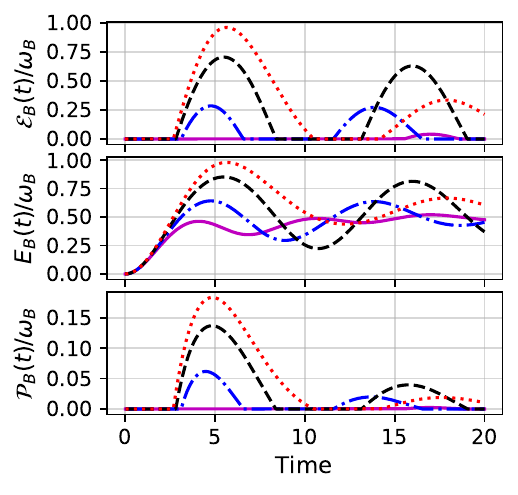}%
		}
		\hfill
		\subfloat[\label{CHARGING_EXEMPLE_1_entangled}]{%
			\includegraphics[width=0.69\columnwidth]{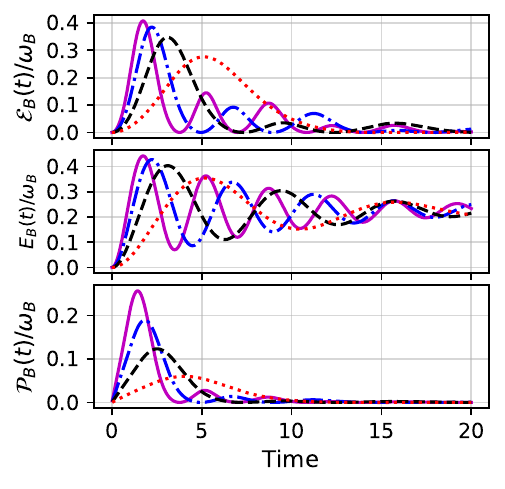}%
		}\hfill
		\subfloat[\label{CHARGING_EXEMPLE_2_entangled}]{%
			\includegraphics[width=0.69\columnwidth]{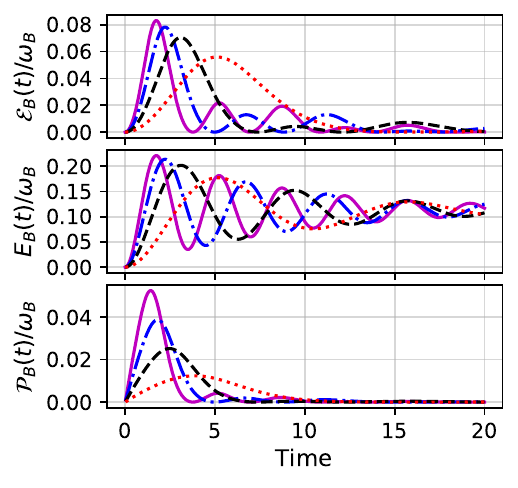}%
		}\hfill
		\subfloat[\label{CHARGING_EXEMPLE_3_entangled}]{%
			\includegraphics[width=0.69\columnwidth]{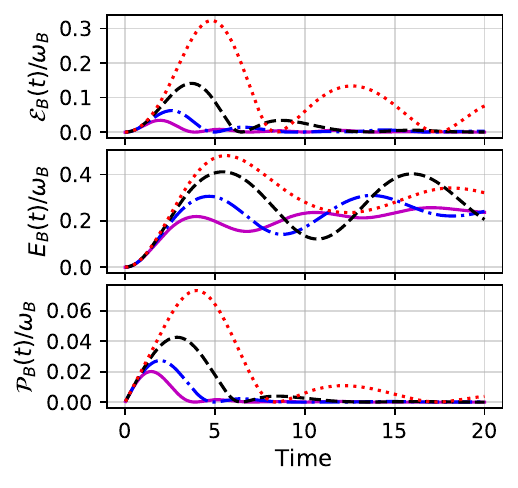}%
		}
		
		\caption{Dynamics of ergotropy, energy, and charging power of the quantum battery versus time, namely  $\frac{\mathcal{E}_B(t)}{\omega_{B}}$, $\frac{E_B(t)}{\omega_{B}}$, and $\frac{\mathcal{P}_B(t)}{\omega_{B}}$, respectively. Panels (a)--(c) correspond to scenarios ($I_a$, $II_a$, $III_a$), while panels (d)--(f) correspond to scenarios ($I_b$, $II_b$, $III_b$). Moreover, $g=0.03,\,0.05,\,0.07,$ and $0.09$ (in units of $\omega_{S_2}$) for red dotted, black dashed, blue dot--dashed and magenta solid curves, respectively.}
		\label{CHARGING}	
	\end{figure*}
	
	In Fig.~(\ref{CHARGING}), we display the dynamics of the ergotropy, stored energy, and charging power of the quantum battery as functions of time $t$. We plot these physical quantities for scenarios ($I_a$, $II_a$, $III_a$) and ($I_b$, $II_b$, $III_b$) in Figs.~(\ref{CHARGING_EXEMPLE_1_010}, \ref{CHARGING_EXEMPLE_2_0110}, \ref{CHARGING_EXEMPLE_3_0110}) and Figs.~(\ref{CHARGING_EXEMPLE_1_entangled}, \ref{CHARGING_EXEMPLE_2_entangled}, \ref{CHARGING_EXEMPLE_3_entangled}), respectively. 
	
	Clearly, one can observe that for ($I_{a,b}$, $II_{a,b}$), the ergotropy, stored energy, and charging power increase over time as the coupling strength $g$ increases. In contrast, for scenario ($III_{a,b}$), these quantities increase as $g$ decreases. A detailed discussion on the impact of the coupling strength on energy extraction is provided in Appendix~\ref{effectofgandk}.  Physically speaking, for ($I_{a,b}$, $II_{a,b}$), the interaction is common between $S_{12}-B$ and $S_{12}-CB$ and according to the interaction Hamiltonians of the scenarios ($I$ and $II$) given in Eqs.~\ref{INT1} and \ref{INT2}. In fact, they bias the following transitions:
	
	\begin{align*}
		\ket{0_{S_1}1_{S_2}0_B} \to \ket{1_{S_1}0_{S_2}1_B}\quad &\text{for Scenario I},\\
		\ket{0_{S_1}1_{S_2}1_C 0_B} \to \ket{1_{S_1}0_{S_2}0_C 1_B}\quad &\text{for Scenario II}.
	\end{align*}
	This implies that the quantum battery is transferred from the passive state $\ket{0_B}$ to the fully active state $\ket{1_B}$, in both scenarios ($I$ and $II$).   In contrast, if $g$ increases, the ergotropy decreases in scenario ($III$). The transition from a non-passive state of quantum battery to a fully passive state occurs immediately via a quantum charger. Here, the charger acts as part of the structured reservoir in the quantum charging setup $RS_{12}C\text{--}B$. Moreover, we observe that the plotted quantities decrease when transitioning from ($I_a$, $II_a$, $III_a$) to ($I_b$, $II_b$, $III_b$), which means that quantum battery transfers to the passive state in ($I_a$, $II_a$, $III_a$) more efficiently than in ($I_b$, $II_b$, $III_b$). In the next section, we shall analyze why the battery transfers to the passive state in ($I_a$, $II_a$, $III_a$) more efficiently than in ($I_b$, $II_b$, $III_b$), where we study the effect of population and coherence on the charging of the quantum battery over time.

	\subsection{Effects of coherence and population on the charging process}
	\label{sec:Effects of coherence and population on the charging process}
	
	In Sec.~\ref{sec:Charging Process}, we demonstrated numerically the ability of each scenario to extract work over time, without providing any information on the effect of coherence, correlation and population on the behavior of the ergotropy. However, to address this gap and to provide a clearer theoretical understanding, we analyze in this part the effect of energy of coherence and energy of populations and correlations on the charging process in the scenarios ($I_a$, $II_a$, and $III_a$) and ($I_b$, $II_b$, and $III_b$).

	As seen in Eqs.~\ref{ERGO_COH} and \ref{ERGO_POP}, the total ergotropy of quantum battery is composed of two parts, namely the ergotropy $\mathcal{E}_{B}^{C}(t)$ due to the coherence energy and $\mathcal{E}_{B}^{P}(t)$, which reflects the exchange of excitation (population inversion). Physically, this means that we can charge the quantum battery $B$ using two methods, that is, coherence transfer and excitation exchange between the subsystems $S_{12}=S_1 \otimes S_2$, charger $C$ and battery $B$ over time. Then, in Fig.~\ref{ERGOTROPY}, we examine $\mathcal{E}_B^{P}(t)$ and $\mathcal{E}_B^{C}(t)$ numerically for the two examples $S_a$ and $S_b$ in Figs.~(\ref{ERGO_EXEMPLE_1_010}, \ref{ERGO_EXEMPLE_2_0110}, \ref{ERGO_EXEMPLE_3_0110}) and Figs.~(\ref{ERGO_EXEMPLE_1_entangled}, \ref{ERGO_EXEMPLE_2_entangled}, \ref{ERGO_EXEMPLE_3_entangled}), respectively.   
	The coherent contribution initially increases from zero and subsequently exhibits damped oscillations due to the competition between coherent energy exchange and reservoir-induced dissipation.
	
	For the semi-classical state of the total system $S_a$ with the initial state $\hat{\rho}_{S_a}(0)$, we observe that the coherence ergotropy vanishes, namely $\mathcal{E}_B^{C}(t)=0$ for any value of the coupling $g$. However, the population ergotropy $\mathcal{E}_B^{P}(t)$ fluctuates between its maximum and minimum bounds until vanishing for large values of $t$. Physically, this means that the ergotropy is due only to the exchange of excitation between the subsystems over time.\\
	In the presence of coherence in the total system $S_b$, with the initial state $\hat{\rho}_{S_b}(0)$, we observe that the coherence ergotropy $\mathcal{E}_B^{C}(t)$ increases over time for any value of $g$, with oscillatory behavior. In contrast, the population ergotropy $\mathcal{E}_B^{P}(t)$ vanishes over time. Physically, this indicates that work can be extracted from the battery only through coherence during the dynamics.

	\begin{figure*}[htp!]
		
		\subfloat[  \label{ERGO_EXEMPLE_1_010}]{%
			\includegraphics[width=0.69\columnwidth]{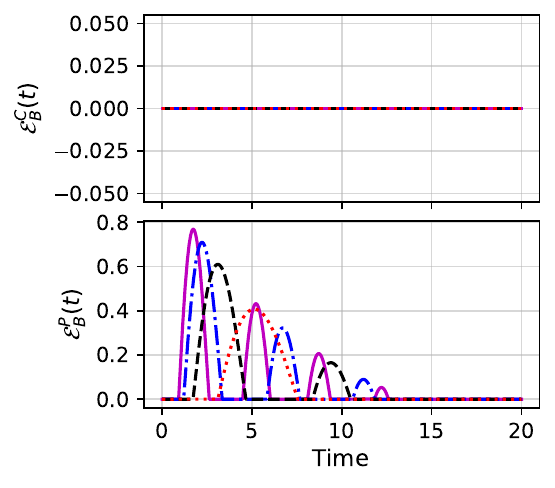}%
		}\hfill
		\subfloat[\label{ERGO_EXEMPLE_2_0110}]{%
			\includegraphics[width=0.69\columnwidth]{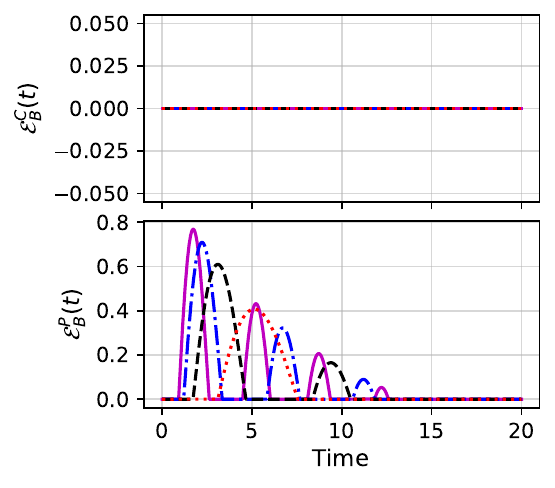}%
		}\hfill
		\subfloat[\label{ERGO_EXEMPLE_3_0110}]{%
			\includegraphics[width=0.69\columnwidth]{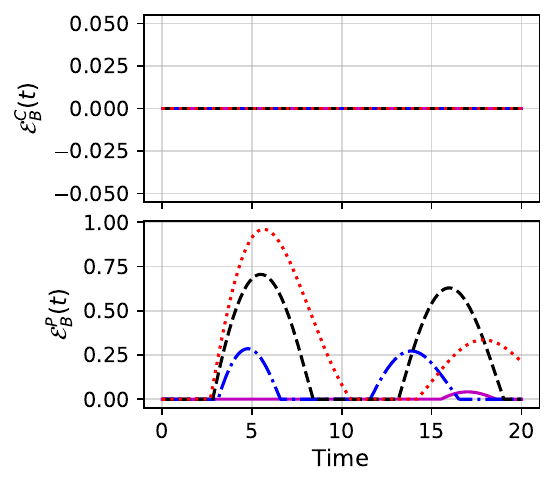}%
		}
		\hfill
		\subfloat[\label{ERGO_EXEMPLE_1_entangled}]{%
			\includegraphics[width=0.69\columnwidth]{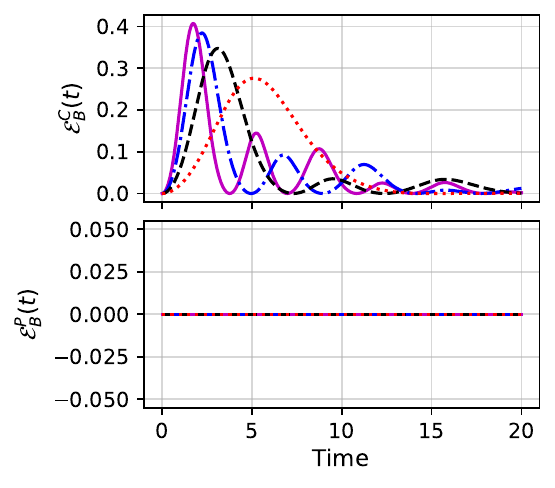}%
		}\hfill
		\subfloat[\label{ERGO_EXEMPLE_2_entangled}]{%
			\includegraphics[width=0.69\columnwidth]{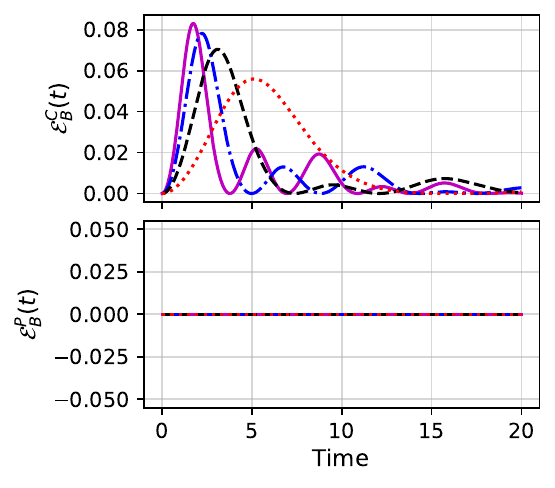}%
		}\hfill
		\subfloat[\label{ERGO_EXEMPLE_3_entangled}]{%
			\includegraphics[width=0.69\columnwidth]{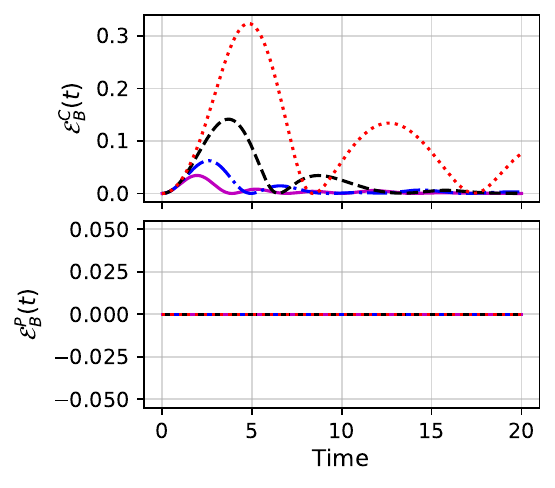}%
		}
		
		\caption{Dynamics of ergotropy of coherence as well as ergotropy of population of the quantum battery, respectively. Moreover, panels (a-c) correspond to scenarios ($I_a$, $II_a$, $III_a$), while panels (d-f) treat scenarios ($I_b$, $II_b$, $III_b$). However, we set $g=0.03,\,0.05,\,0.07,$ and $0.09$ (in units of $\omega_{S_2}$) for red, black, blue and magenta curves, respectively.}	\label{ERGOTROPY}	
	\end{figure*}
	
	Hence, one can conclude that in the case of example $S_a$, the total ergotropy is due only to the population contribution, such that $\mathcal{E}_B(t)=\mathcal{E}_B^{P}(t)$. However, for example, $S_b$, the total ergotropy does not vanish according to coherence only, so $\mathcal{E}_B(t)=\mathcal{E}_B^{C}(t)$. According to Eq.~\ref{RGO_COH_COR_Bounds}, the total ergotropy is bounded by the population ergotropy and the energies of coherence $\mathcal{W}_{coh}(t)$ and correlations $\mathcal{W}_{cor}(t)$. To analyze these effects on the charging process, we plot in Fig.~\ref{Coherence_energy} the resulting ergotropies for examples $S_a$ and $S_b$ in Figs.~(\ref{CORRELATION_energy_EXEMPLE_1_classical}, \ref{CORRELATION_energy_EXEMPLE_2_classical}, \ref{CORRELATION_energy_EXEMPLE_3_classical}) and Figs.~(\ref{CORRELATION_energy_EXEMPLE_1_entangled}, \ref{CORRELATION_energy_EXEMPLE_2_entangled}, \ref{CORRELATION_energy_EXEMPLE_3_entangled}), respectively.
	\begin{figure*}[htp!]
		
		\subfloat[  \label{CORRELATION_energy_EXEMPLE_1_classical}]{%
			\includegraphics[width=0.69\columnwidth]{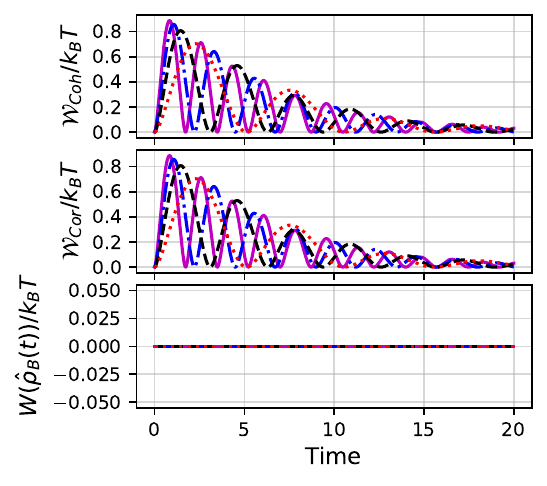}%
		}\hfill
		\subfloat[\label{CORRELATION_energy_EXEMPLE_2_classical}]{%
			\includegraphics[width=0.69\columnwidth]{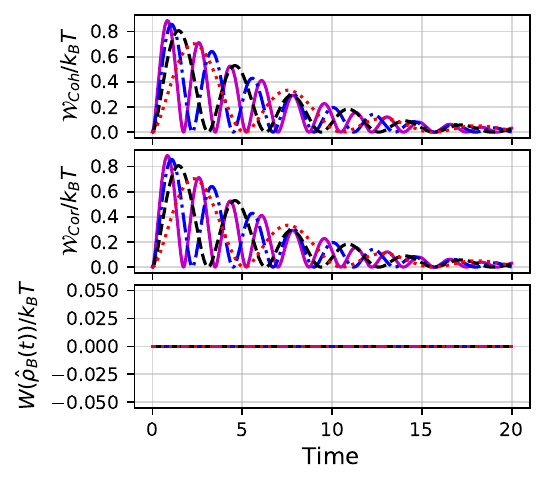}%
		}\hfill
		\subfloat[\label{CORRELATION_energy_EXEMPLE_3_classical}]{%
			\includegraphics[width=0.69\columnwidth]{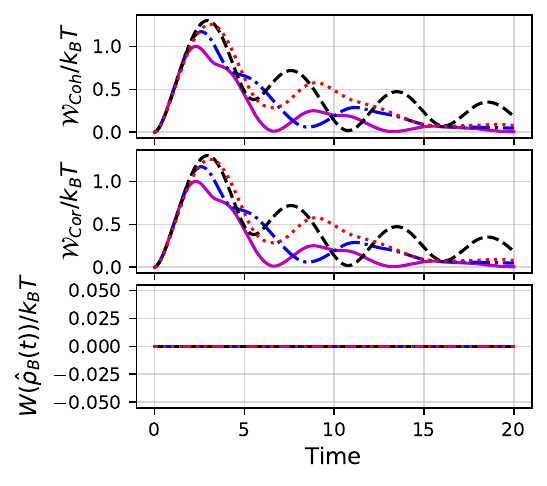}%
		}
		\hfill
		\subfloat[\label{CORRELATION_energy_EXEMPLE_1_entangled}]{%
			\includegraphics[width=0.69\columnwidth]{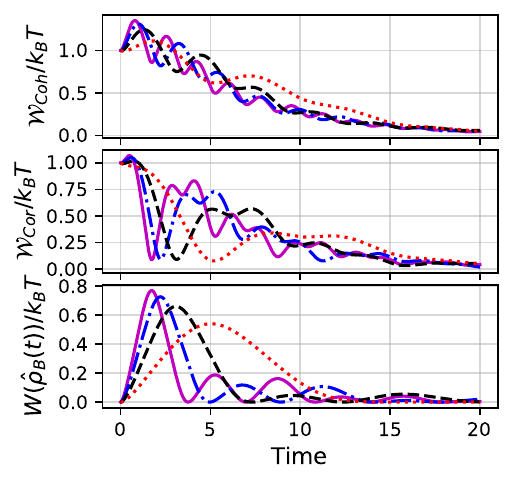}%
		}\hfill
		\subfloat[\label{CORRELATION_energy_EXEMPLE_2_entangled}]{%
			\includegraphics[width=0.69\columnwidth]{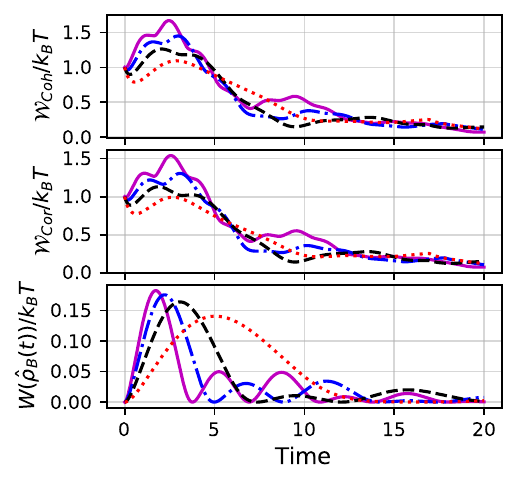}%
		}\hfill
		\subfloat[\label{CORRELATION_energy_EXEMPLE_3_entangled}]{%
			\includegraphics[width=0.69\columnwidth]{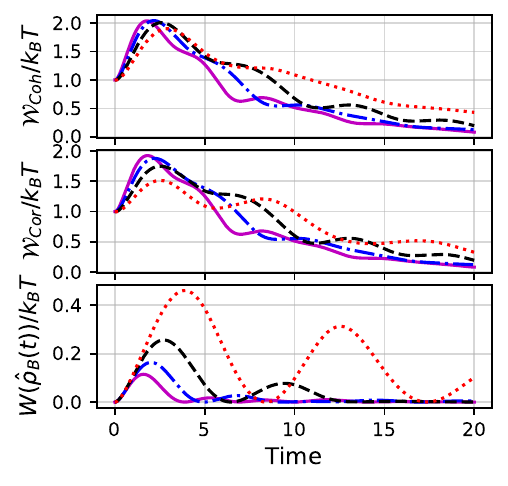}%
		}

		\caption{Dynamics of coherence contribution $W_{\mathrm{coh}}/k_BT$, correlation contribution $W_{\mathrm{cor}}/k_BT$ and free energy 
				of coherence of quantum battery $W(\rho_B)/k_BT$. Panels (a)--(c) correspond 
				to scenarios $(I_a,II_a,III_a)$. While, panels (d)--(f) correspond to 
				$(I_b,II_b,III_b)$.} 	\label{Coherence_energy}	
	\end{figure*}
	
	For the example of the classical state of the total system, namely $S_a$ in Figs.~(\ref{CORRELATION_energy_EXEMPLE_1_classical}, \ref{CORRELATION_energy_EXEMPLE_2_classical}, \ref{CORRELATION_energy_EXEMPLE_3_classical}), we observe that for any value of the coupling $g$, the energy stored by means of battery's quantum coherence is vanished, namely $W(\hat{\rho}_B(t)) = 0$, for any value of $t$. Physically, this means that no energy is stored in coherence. Also, this implies that the coherence and correlation energies are equal, $\mathcal{W}_{\mathrm{coh}}(t) = \mathcal{W}_{\mathrm{cor}}(t)$. Therefore, the bound on the energy stored in coherence given in Eq.~\ref{COH_COR_Bound} is not satisfied.\\
	Importantly, from Figs.~(\ref{ERGO_EXEMPLE_1_010}, \ref{ERGO_EXEMPLE_2_0110}, \ref{ERGO_EXEMPLE_3_0110}), the coherence and population ergotropy for the example $S_a$ satisfy $\mathcal{E}_B^{P}(t) = \mathcal{E}_B(t)$ and $\mathcal{E}_B^{C}(t) = 0$, which is also consistent with the bound in Eq.~\ref{RGO_COH_COR_Bounds}, where the battery ergotropy is bounded as
	$$\mathcal{E}_B^{P}(t) \leq \mathcal{E}_B(t) \leq \mathcal{E}_B^{P}(t).$$
	Physically speaking, the absence of coherence energy storage is due to the consumption of coherence energy as correlations to sustain the memory effects provided by the structured reservoir. However, the energy is stored only in the population of the quantum battery under the probability inversion imposed by the interaction Hamiltonian in each scenario.
	
	In the case of coherence in the total system $S_b$, the results are reported in Figs.~(\ref{CORRELATION_energy_EXEMPLE_1_entangled}, \ref{CORRELATION_energy_EXEMPLE_2_entangled}, \ref{CORRELATION_energy_EXEMPLE_3_entangled}). For any value of the coupling $g$, the free energy of coherence of the quantum battery vanishes initially and subsequently increases over time, exhibiting oscillatory behavior. This behavior is due to the fact that the initial correlation energy is equal to the initial coherence energy, i.e, $\mathcal{W}_{\mathrm{coh}}(0) = \mathcal{W}_{\mathrm{cor}}(0)$. While the energy stored in coherence becomes higher than that consumed by correlations over time. Physically, the presence of coherence initially and the initial correlations of the structured qubits in the quantum reservoir enhance the transfer of coherence between the subsystems over time.

	\begin{figure*}[htp!]
		
		\subfloat[  \label{ERGOvsCOR_QUANTUM_EX1}]{%
			\includegraphics[width=0.58\columnwidth]{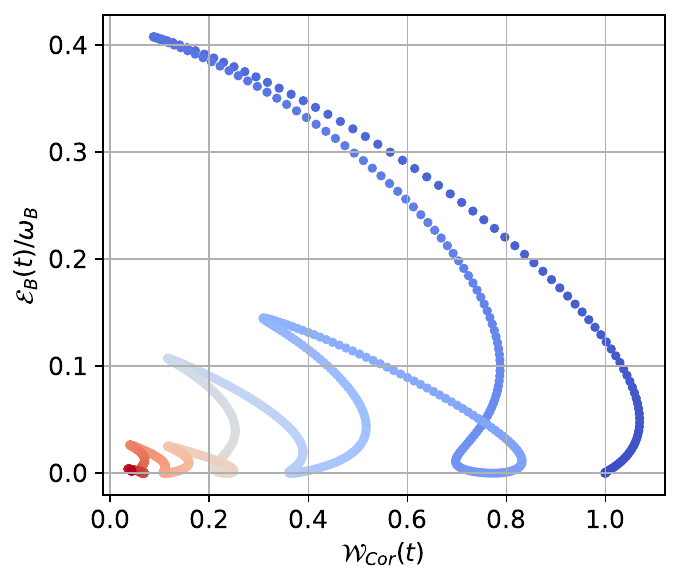}%
		}\hfill
		\subfloat[\label{ERGOvsCOR_QUANTUM_EX2}]{%
			\includegraphics[width=0.59\columnwidth]{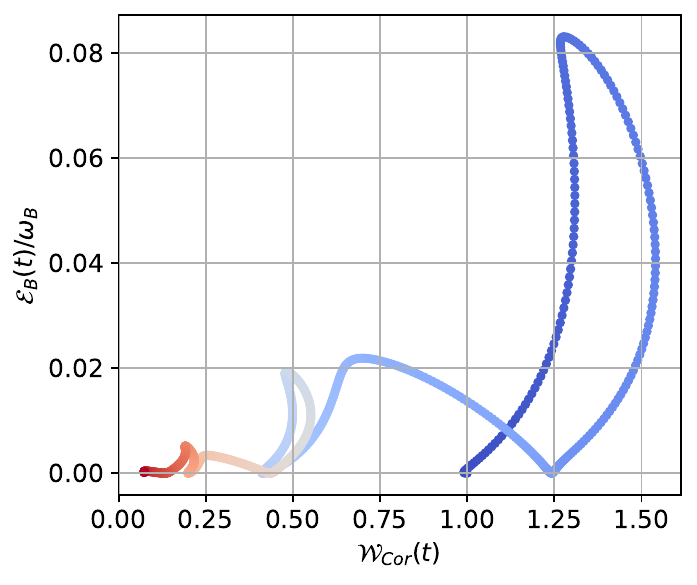}%
		}\hfill
		\subfloat[\label{ERGOvsCOR_QUANTUM_EX3}]{%
			\includegraphics[width=0.69\columnwidth]{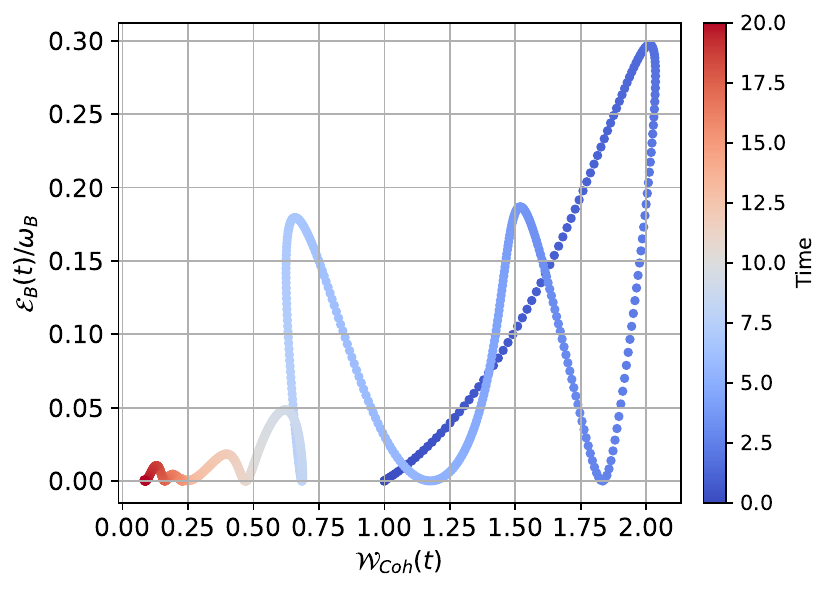}%
		}
		\hfill
		\subfloat[\label{ERGOvsCOH_QUANTUM_EX1}]{%
			\includegraphics[width=0.58\columnwidth]{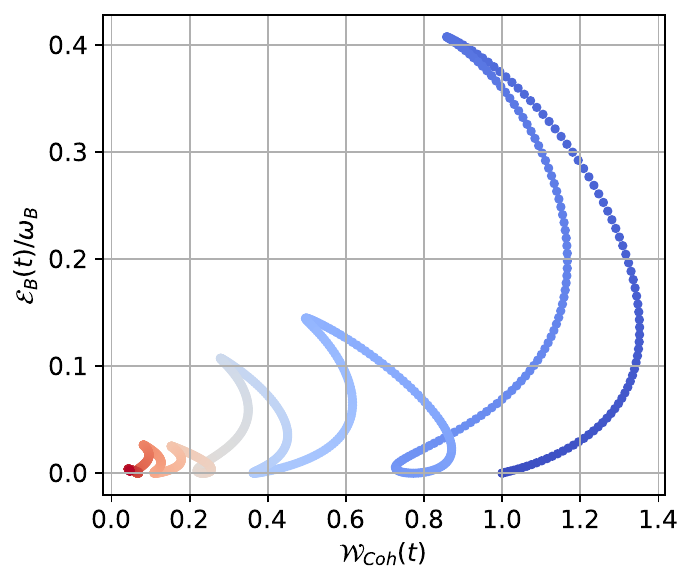}%
		}\hfill
		\subfloat[\label{ERGOvsCOH_QUANTUM_EX2}]{%
			\includegraphics[width=0.59\columnwidth]{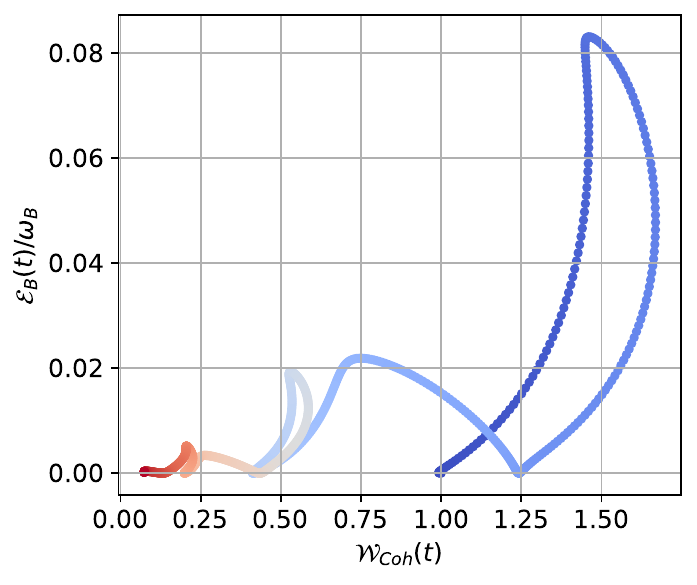}%
		}\hfill
		\subfloat[\label{ERGOvsCOH_QUANTUM_EX3}]{%
			\includegraphics[width=0.69\columnwidth]{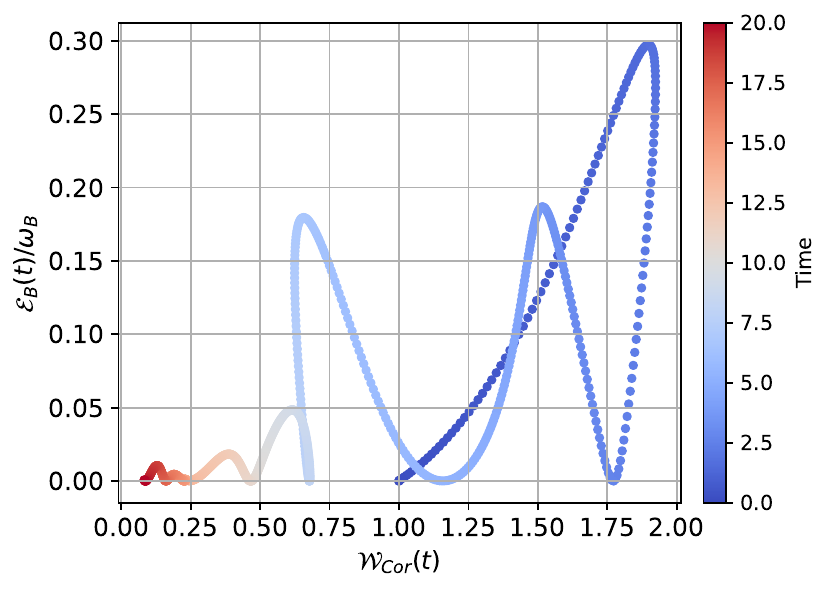}%
		}

\caption{
Parametric plots of the battery ergotropy $\mathcal{E}_B(t)/\omega_B$ 
	as a function of the correlation contribution $W_{\mathrm{cor}}(t)$ 
	and coherence contribution $W_{\mathrm{coh}}(t)$ for the coherent 
	initial-state scenarios $(I_b,II_b,III_b)$. Panels (a), (b), and (f) 
	show $\mathcal{E}_B(t)/\omega_B$ as a function of $W_{\mathrm{cor}}(t)$ 
	for scenarios $I_b$, $II_b$, and $III_b$, respectively. Panels (d), 
	(e), and (c) show $\mathcal{E}_B(t)/\omega_B$ as a function of 
	$W_{\mathrm{coh}}(t)$ for scenarios $I_b$, $II_b$, and $III_b$, 
	respectively. The color gradient indicates the time evolution. 
	We set $g=0.09$ and $k=0.03$ in units of $\omega_{S_2}$.
}

	\label{PAP}	
	\end{figure*}
	In Fig.~\ref{PAP}, we analyze the impact of correlation energy $\mathcal{W}_{\mathrm{cor}}(t)$ and coherence energy $\mathcal{W}_{\mathrm{coh}}(t)$ on the ergotropy of the quantum battery, namely $\mathcal{E}_B(t)/\omega_B$. This is illustrated in Figs.~\ref{PAP}(a--b) and \ref{PAP}(d--f) for the scenarios ($I_b$, $II_b$, $III_b$). Indeed, we focus on these scenarios rather than ($I_a$, $II_a$, $III_a$) because for an initially incoherent state the energy stored in the coherence of the quantum battery satisfies $W(\hat{\rho}_B(t))/K_B T = 0$, implying $\mathcal{W}_{\mathrm{coh}}(t) = \mathcal{W}_{\mathrm{cor}}(t) \geq 0$. Physically, this indicates that although the interaction between subsystems generates global coherence and correlations. The latter acts as an information reservoir for backflow (memory effects) rather than as a resource for work extraction.
	
	In contrast, for an initially coherent state, Fig.~\ref{PAP} highlights that after an initial transient regime in which correlations build up, where coherence contribution dominates, namely $\mathcal{W}_{\mathrm{coh}}(t) > \mathcal{W}_{\mathrm{cor}}(t)$. Consequently, this leads to a positive ergotropy of the quantum battery, that is, $\mathcal{E}_B(t)/\omega_B > 0$. Physically, this demonstrates that correlations play a dual role, i.e, they consume initially coherence to establish correlations among subsystems, and subsequently act as a mediator that facilitates coherence transfer from the structured reservoir to the battery. This correlation-assisted redistribution of coherence enhances the charging performance by enabling more efficient and collective energy transfer processes.
\section{Discussions}\label{DISCU}

Our fundamental results show that the correlations between the subsystems play an important role in the case of structured reservoirs. Moreover, according to the scenarios presented in this work, the battery can be charged autonomously without any external source. Importantly, we showed that the bound on the ergotropy in Eq.\ref{RGO_COH_COR_Bounds} is not universal and it is limited by the inequality in Eq.\ref{CONDITION_VALIDITY}. Compared to the more general bound on ergotropy, which describes the bound on coherence ergotropy with respect to the thermal state in \cite{FINALE}, our bound gives a description of how the contributions of global coherence and correlations between the subsystems affect ergotropy production in the quantum battery. On the other hand, the contribution of the energy stored in correlations, namely $\mathcal{W}_{cor}(t)$ highlights how correlations contribute to ergotropy production over time.

The role of the ancillas $S_1$ and $S_2$ of the structured reservoirs still extremely important for redistributing coherence and correlations between the subsystems over time. As mentioned before, the dissipator term in Eq.\ref{LL} is described by two contributions: one is due to the interaction Hamiltonian between the subsystem qubits $\hat{H}_{Int_\alpha}$. While, the other contribution is linked to the thermal dissipation induced by the reservoirs $R_1$ and $R_2$ acting on $S_1$ and $S_2$, respectively. The battery stored-energy flow, denoted by $J_B(t)= \frac{d}{dt}E_B(t)$ is composed into two contributions as follows (see Appendix \ref{app:current_decomposition})
\begin{eqnarray*}
		J_B(t)&=&J_{cor}^B(t) + J_{loc}^B(t),\\
		J_{loc}^B(t)&=& -iTr\biggl\{H_B\left[\hat{H}_{I n t_\alpha}, \hat{\rho}_{S_1}(t) \otimes \hat{\rho}_{S_2}(t) \otimes \hat{\rho}_C(t) \otimes \hat{\rho}_B(t)\right]\biggl\},\\
		J_{cor}^B(t)&=& -iTr\biggl\{\hat{H}_B\left[H_{I n t_\alpha}, \hat{\rho}_{cor}(t) \right]\biggl\}, 
\end{eqnarray*}
where $H_B \equiv \mathbb{I}_{S_1 S_2 C}\otimes \hat{H}_B$. However, $J_{loc}^B(t)$ denotes the local contribution to the energy stored in the battery, while $J_{cor}^B(t)$ represents the contribution of correlations to the energy current of the quantum battery. Therefore, this expression provides a quantitative measure of the correlation-induced contribution to the energy flow into the battery. Consequently, correlations are not introduced only as a qualitative interpretation. In fact, they enter explicitly through the non-product part $\hat{\rho}_{cor}(t)$ of the total state. Moreover, note that we do not identify $J_{cor}^B(t)$ and $W_{\mathrm{cor}}(t)$ with the ergotropy itself. But, we used them to understand the effect of correlations and how the correlation contribution affects and bounds the work extracted from the quantum battery $B$.

Moreover, since the reservoir temperature is positive, the thermal state remains passive, then no ergotropy can be extracted directly from a single thermal reservoir, as discussed in \cite{FINAL1}. In other words, although the battery can store thermal energy, the extractable work remains zero. In our model, the situation is fundamentally different because the charging process is mediated by two thermal reservoirs, namely $R_1$ and $R_2$, through a structured reservoir architecture involving the auxiliary qubits $S_1$ and $S_2$, as well as the charger $C$. In general, by controlling the temperatures of the two reservoirs, one may generate a virtual temperature and induce population inversion, which can lead to nonzero ergotropy stored in the battery. However, throughout the present work we consider the symmetric case, that is, $T_1=T_2$. Under this condition, a direct coupling of the battery to the thermal reservoirs will not generate any population inversion and therefore no ergotropy is produced. Then, the battery would simply relax toward a passive thermal state. Therefore, if the battery is initially prepared in a passive state, one can obtain
	$$
	\mathcal{E}_B(t)=0,
	\qquad \forall t .
	$$

If, instead, the battery is initially prepared in an active state, its ergotropy vanishes at equilibrium, namely
	$$
	\mathcal{E}_B(t_{\mathrm{equilibrium}})=0,
	\qquad t_{\mathrm{equilibrium}}\to +\infty .
	$$
	In contrast, when the structured reservoir qubits ($S_1$) and ($S_2$) and the charger ($C$) are included, the battery ergotropy becomes nonzero and strongly depends on the considered interaction scenario. This enhancement does not originate from direct thermal excitation, but rather from the coherent transfer of energy and coherence through the structured reservoir architecture, together with the temporary storage of energy in the correlations established among the subsystems. Consequently, the structured reservoir acts as a mediator that enables the generation and redistribution of useful work resources, leading to a significant enhancement of the battery charging performance compared to the direct reservoir--battery coupling case.
	\begin{figure*}[htp!]
		\centering
		\subfloat[\label{fig:COMP_A}]{%
			\includegraphics[width=0.95\textwidth]{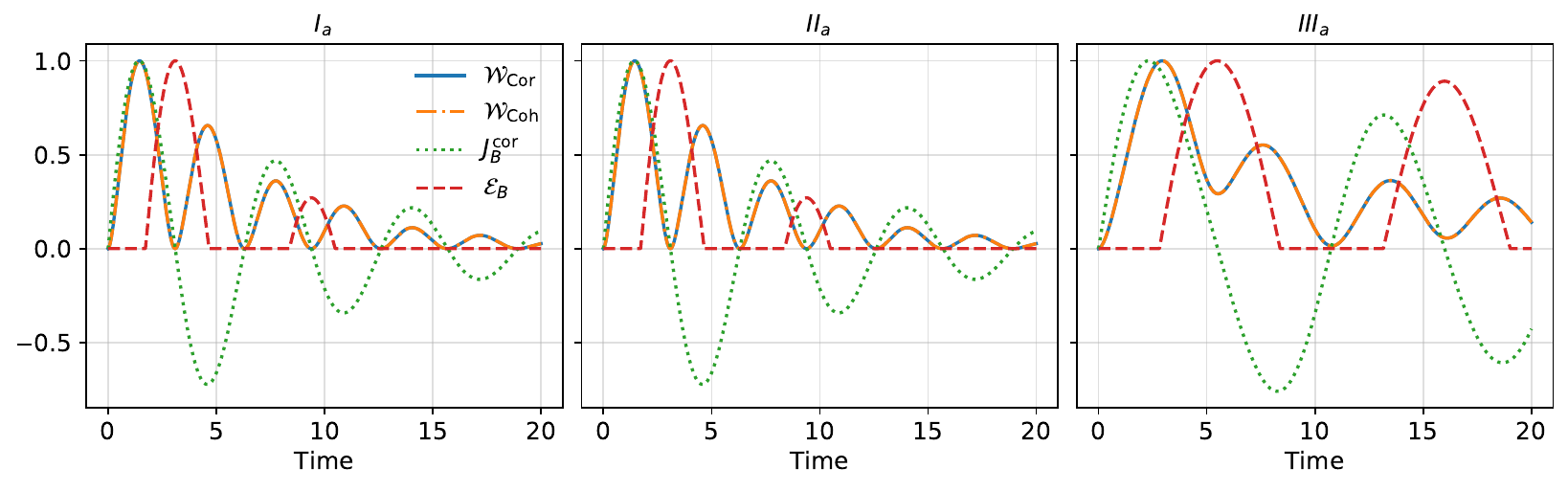}%
		}
		
		\vspace{0.3cm}
		
		\subfloat[\label{fig:COMP_B}]{%
			\includegraphics[width=0.95\textwidth]{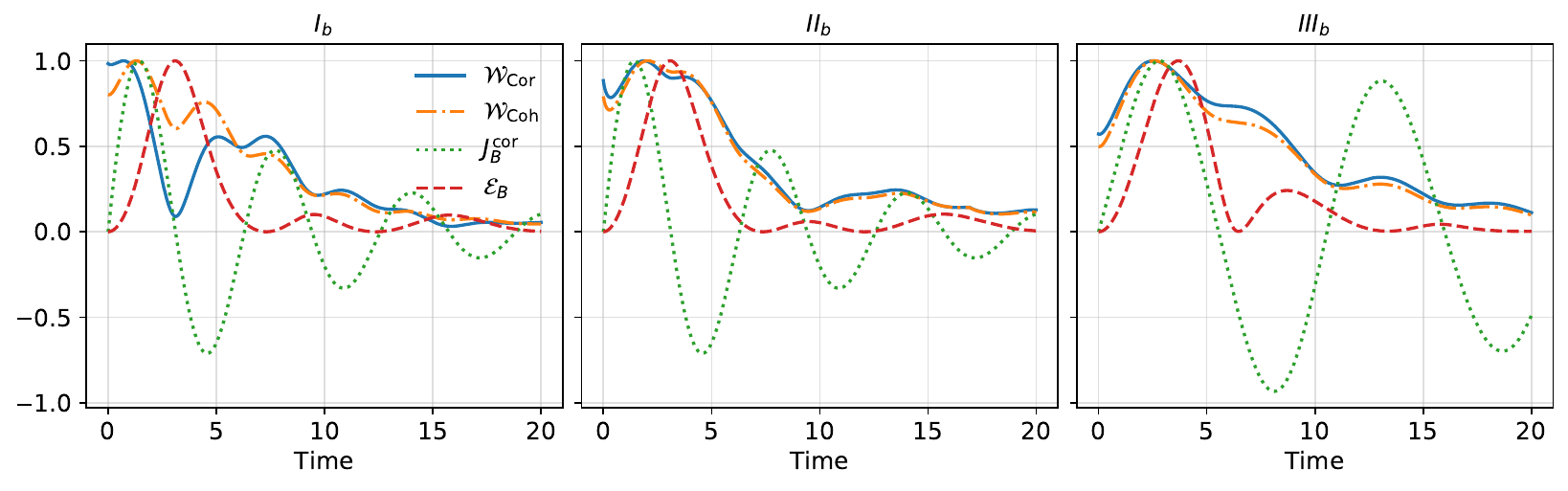}%
		}
		
		\caption{Time evolution of the normalized correlation energy, normalized coherence energy, normalized correlation-induced energy current, and normalized battery ergotropy for the three charging scenarios. Panel (a) corresponds to scenarios $I_a$, $II_a$, and $III_a$, while panel (b) corresponds to scenarios $I_b$, $II_b$, and $III_b$. The plotted quantities are respectively normalized as
				$\mathcal{W}_{\mathrm{cor}}(t)/\max_t|\mathcal{W}_{\mathrm{cor}}(t)|$,
				$\mathcal{W}_{\mathrm{coh}}(t)/\max_t|\mathcal{W}_{\mathrm{coh}}(t)|$,
				$J_B^{\mathrm{cor}}(t)/\max_t|J_B^{\mathrm{cor}}(t)|$, and
				$\mathcal{E}_B(t)/\max_t|\mathcal{E}_B(t)|$.
				Besides, we set $g=0.05$ and $k=0.03$ in units of $\omega_{S_2}$.}
		\label{fig:COMP_AB}
	\end{figure*}

Figure~\ref{fig:COMP_AB} shows the normalized time evolution of 
	$\mathcal{W}_{\mathrm{cor}}(t)$, $\mathcal{W}_{\mathrm{coh}}(t)$, 
	$J_B^{\mathrm{cor}}(t)$, and $\mathcal{E}_B(t)$ for the three charging scenarios. 
	Since all quantities are normalized by their maximum absolute values, the figure is used only to compare their temporal behavior and not their absolute amplitudes.

For the initial states $S_a$, shown in panel~(a), the dynamics is mainly driven by population transfer. In scenarios $I_a$ and $II_a$, the curves exhibit very similar oscillatory behavior, indicating that both scenarios generate an effective resonant transfer channel toward the battery. Since the battery remains incoherent for the initial state $S_a$, its free energy of coherence vanishes, such that
	\begin{equation}
		\mathcal{W}_{\mathrm{coh}}(t)
		=
		\mathcal{W}_{\mathrm{cor}}(t),
	\end{equation}
	throughout the evolution. Consequently, the battery ergotropy originates entirely from the population contribution,
	$\mathcal{E}_B(t)=\mathcal{E}_B^P(t)$, rather than from coherence. In scenario $III_a$, the dynamics differs because the charging process occurs through a sequential mechanism involving the charger. The more pronounced oscillations of $J_B^{\mathrm{cor}}(t)$ indicate a stronger exchange of correlations among the subsystems during the charging process. Nevertheless, the battery ergotropy remains population-driven, while the charger mediates the redistribution and transfer of energy toward the battery.

For the coherent initial states $S_b$, shown in panel (b), the initial coherence in the structured reservoir and charger modifies the charging dynamics. Compared with panel (a), the oscillations become smoother and more persistent. This indicates that the initial coherence contributes to maintaining the energy exchange between the subsystems over a longer time. In particular, the behavior of $\mathcal{E}_B(t)$ follows the balance between $\mathcal{W}_{\mathrm{coh}}(t)$ and $\mathcal{W}_{\mathrm{cor}}(t)$, showing that the extractable work is enhanced when the coherence contribution is not completely consumed by correlations.

Overall, these results show that correlations do not simply reduce the charging efficiency. Instead, they mediate the redistribution of coherence between the structured reservoir, charger, and battery. When the coherence contribution exceeds the correlation contribution, the battery can store useful extractable work. Therefore, the normalized plots provide numerical evidence that the charging process is governed by the competition between coherence transfer and correlation exchange.

	\section{Conclusion}	
	\label{sec:Conclusion}
	We have developed a theoretical framework to investigate the effect of structured reservoirs on the charging of a quantum battery through three distinct interaction scenarios. Our analysis identifies coherence, population, and correlation exchange as key resources governing the charging dynamics and the extractable work. Moreover, we derived bounds on the ergotropy in terms of coherence and correlations for each scenario. In particular, we showed that the ergotropy is bounded from below by the population contribution and from above by the sum of the population ergotropy and the coherence energy stored in the quantum battery, reduced by the energy associated with correlations between the subsystems. In this regard, we conclude that coherence serves as a valuable resource for enhancing the charging process, while correlations play a crucial role in preserving information during the non-unitary evolution induced by the local master equation.
	
	Importantly, we demonstrated that for incoherent initial states, the ergotropy originates solely from population contributions. In contrast, coherent initial states enable additional contributions arising from coherence transfer and correlation backflow. We further derived analytical bounds on the ergotropy in terms of the free energy of coherence and showed their consistency with numerical simulations. In addition, our results reveal that the interplay between structured reservoirs and autonomous charging dynamics provides a mechanism for enhancing quantum battery performance without the need for external work. In this regard, our approach extends previous studies by explicitly incorporating the effects of structured environments and offers a pathway toward experimental realization in superconducting qubit platforms.
	
	Although our analysis focuses on bosonic thermal reservoirs, where the underlying mechanism relies on correlation-assisted energy exchange induced by structured couplings. While different bath statistics may modify transition rates, where they are not expected to alter the qualitative behavior. However, non-thermal reservoirs, such as squeezed baths, may further enhance coherence generation and thus represent an interesting direction for future research.
	
	Overall, our work generalizes previous results on coherence-driven and reservoir-assisted quantum batteries by explicitly incorporating structured reservoirs, memory effects, and multi-qubit Hilbert-space structures. It also provides experimentally relevant conditions for implementing autonomous, resource-enhanced quantum battery charging in superconducting qubit platforms.
	

	\acknowledgments
	A.~K, acknowledges CNRST-Morocco support for this research within the Program " PhD-ASsociate Scholarship – PASS".
	
	\section*{Declaration of Interest}
	The authors declare that they have no conflict of interest.
	
	\section*{Data availability statement}
	No data statement is available.
	
	

\appendix
\section{Demonstration of the ergotropy bound for a quantum battery}
\label{Ergotropy_bound}

\subsection{Theoretical description of Eq.~\ref{COHERENCE BOUND}}
\label{Theoretical}
\subsubsection{Theoretical framework}

The ergotropy of a quantum battery can be decomposed into population and coherence contributions as~\cite{MODEL10,FINAL1,FINAL2}
	\[
	\mathcal{E}_B = \mathcal{E}_B^{P} + \mathcal{E}_B^{C}.
	\]
	The coherent contribution is defined as the difference between the ergotropy of the full state and that of the dephased state. In fact, it is defined as: 
	\[
	\mathcal{E}_B^{C} = \mathcal{E}_B - \mathcal{E}_B^{P}.
	\]

The total ergotropy satisfies the following well-known free-energy bound:
	\[
	\mathcal{E}_B \le F(\hat{\rho}_B) - F(\hat{\rho}_B^{\mathrm{th}}).
	\]
	Similarly, the population contribution satisfies
	\[
	\mathcal{E}_B^{P} \le F(\bar{\rho}_B) - F(\hat{\rho}_B^{\mathrm{th}}),
	\]
	where $\bar{\rho}_B$ and $\hat{\rho}_B^{\mathrm{th}}$ denote the fully dephased state in the energy basis and the thermal state, respectively. Subtracting these inequalities yields the following inequality:
	\[
	\mathcal{E}_B^{C} \le F(\hat{\rho}_B) - F(\bar{\rho}_B).
	\]

The right-hand side of the above equation corresponds to the free-energy contribution associated with coherence, which can be written as
	\[
	W(\hat{\rho}_B) =F(\hat{\rho}_B) - F(\bar{\rho}_B) = k_B T\, C(\hat{\rho}_B),
	\]
	where $C(\hat{\rho}_B)$ is the relative entropy of coherence in the energy eigenbasis. This leads to the following bound:
	\[
	0 \le \mathcal{E}_B^{C} \le k_B T\, C(\hat{\rho}_B).
	\]
	
This bound is a physically motivated upper bound, consistent with the more general results derived in~\cite{FINALE}, which incorporate both coherence and population contributions. The present expression corresponds specifically to the coherence component of these bounds applied to the battery subsystem. Moreover, the bound in Eq.~\eqref{COHERENCE BOUND} is not universal and depends on the following gap quantities, namely $\Delta_{\rm tot}(t)$ and $\Delta_{\rm pop}(t)$ 
	\begin{align}
		\Delta_{\rm tot}(t) &= F(\hat{\rho}_B) - F(\hat{\rho}_B^{\mathrm{th}}) - \mathcal{E}_B, \nonumber\\
		\Delta_{\rm pop}(t) &= F(\bar{\rho}_B) - F(\hat{\rho}_B^{\mathrm{th}}) - \mathcal{E}_B^{P}. \nonumber
	\end{align}
	The subtraction is valid only under the additional condition
	\begin{equation}
		\label{CONDITION_VALIDITY}
		\Delta_{\rm tot}(t) \geq \Delta_{\rm pop}(t).
	\end{equation}
	When this condition is not fulfilled, the quantity $F(\hat{\rho}_B) - F(\bar{\rho}_B)$ should be interpreted as the coherent contribution to the nonequilibrium free energy, rather than as a rigorous upper bound for $\mathcal{E}_B^C(t)$.

\subsubsection{Numerical analysis of the diagnostic gap}

Let us consider the scenarios $(I_a,II_a,III_a)$ and $(I_b,II_b,III_b)$, which are described by the initial states $\hat{\rho}_{S_a}$ and $\hat{\rho}_{S_b}$, respectively. In Fig.~\ref{fig:MANIP_AB}, we numerically analyze the validity of the diagnostic gap described by Eq.~\eqref{CONDITION_VALIDITY}. Mathematically, the diagnostic gap is given by
	\begin{equation}
		\Delta_{\rm tot}(t)-\Delta_{\rm pop}(t) \geq 0.
	\end{equation}
	This condition describes the validity of the bound
	$
	0 \le \mathcal{E}_B^{C} \le k_B T, C(\hat{\rho}_B)
	$
	over time.

For the scenarios $(I_a,II_a,III_a)$, the validity fraction of the diagnostic gap is $100\%$ for all values of the coupling. This is because the ergotropy of the quantum battery is entirely due to the population contribution, since the initial state is incoherent. Therefore, the bound
	$
	0 \le \mathcal{E}_B^{C} \le k_B T, C(\hat{\rho}_B)
	$
	is valid for all times $t\in[0,20]$.

For the scenarios $(I_b,II_b,III_b)$, the validity of the diagnostic bound depends on the coupling strength $g$ and on the scenario under study, due to the fluctuations and the exchange of coherence and correlations between the subsystems.

Physically, this shows that our bound is not universal and depends on the initial state, the coupling strength, and the scenario under study. This is important for highlighting the contribution of coherence and correlations between the subsystems to the ergotropy of the quantum battery.

\begin{figure}[htp!]
	\centering
	\subfloat[\label{fig:MANIP_A}]{%
		\includegraphics[width=0.4\textwidth]{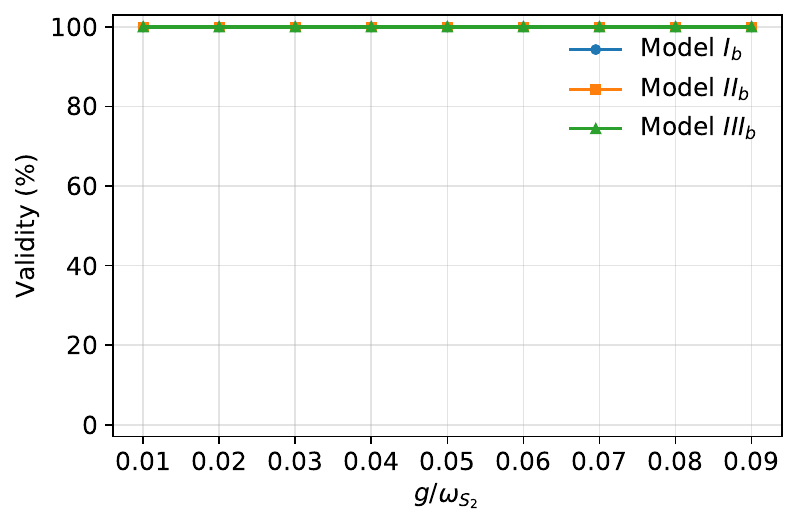}%
	}
	\hfill
	\subfloat[\label{fig:MANIP_B}]{%
		\includegraphics[width=0.4\textwidth]{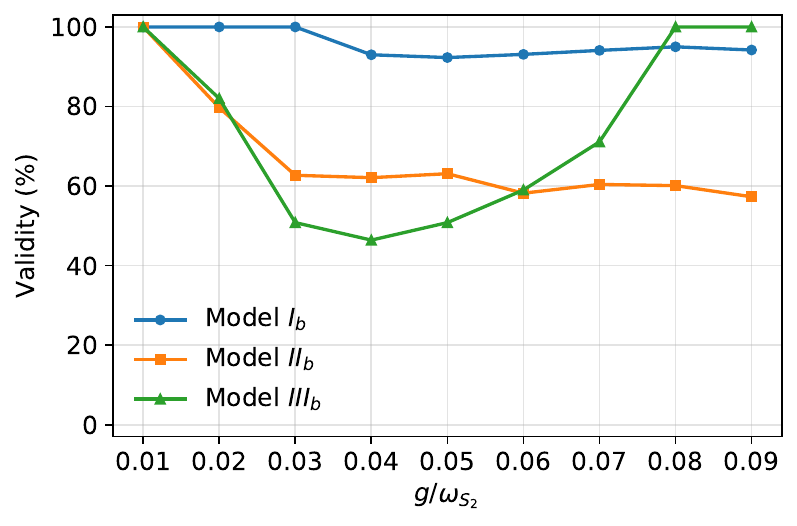}%
	}
	
	\caption{Validity fraction of the gap diagnostic
	($\Delta_{\rm tot}(t)-\Delta_{\rm pop}(t)$) as a function of the normalized coupling strength
	($g/\omega_{S_2}$) for the three interaction scenarios.
	For each value of (g), the validity is computed as the percentage of sampled time points in the interval
	($t\in[0,20]$) satisfying
	($\Delta_{\rm tot}(t)-\Delta_{\rm pop}(t)\geq 0$).
	This condition determines the time intervals over which the coherent-ergotropy inequality
	($0 \le \mathcal{E}_B^{C} \le k_B T, C(\hat{\rho}_B)$) can be consistently evaluated.
	Panel (a) corresponds to scenarios $I_a$, $II_a$, and $III_a$, while panel (b) corresponds to scenarios $I_b$, $II_b$, and $III_b$. We set $k=0.03$ in units of $\omega_{S_2}$.
	}
	\label{fig:MANIP_AB}
\end{figure}
	\subsection{Numerical analysis of Eq.~\ref{COH_COR_Bound}}
	
	\label{Numerical33}
	Equation~\ref{COH_COR_Bound} describes two distinct roles of correlation. In the case of $\mathcal{W}_{\mathrm{coh}} > \mathcal{W}_{\mathrm{cor}}$, the amount of correlation acts as a mediator for coherence transfer between subsystems and enhances work extraction from coherence. In contrast, when $\mathcal{W}_{\mathrm{coh}} = \mathcal{W}_{\mathrm{cor}}$, correlation plays the role of an information reservoir, ensuring memory effects with all coherence converted into correlation.
	
	Numerically, we focus on two cases:
		\begin{itemize}
			\item If $\mathcal{W}_{\mathrm{coh}}(t) > \mathcal{W}_{\mathrm{cor}}(t)$, work can be extracted from coherence, i.e., $\mathcal{E}_B^C(t) > 0$.
			\item If $\mathcal{W}_{\mathrm{coh}}(t) = \mathcal{W}_{\mathrm{cor}}(t)$, no work can be extracted from coherence, i.e., $\mathcal{E}_B^C(t) = 0$.
	\end{itemize}
	
	For this purpose, we first consider scenario $I$, denoted by $I_c$, where the total system $S$ is initially prepared in the state
		\begin{align}
			\hat{\rho}_{S_c}(0) &= \ket{0_{S_1} 1_{S_2} \psi_B}\bra{0_{S_1} 1_{S_2} \psi_B},  \\
			\ket{\psi_B} &= \frac{1}{\sqrt{2}} \left( \ket{0_B} + \ket{1_B} \right). \nonumber
	\end{align}
	
	For scenarios $II$ and $III$, denoted by $II_c$ and $III_c$, the total system $S$ is initially prepared in the state
		\begin{align}
			\hat{\rho}_{S_c}(0) = \ket{0_{S_1} 1_{S_2} 1_C \psi_B}\bra{0_{S_1} 1_{S_2} 1_C \psi_B}.
	\end{align}
	
	In all scenarios ($I_c$, $II_c$, $III_c$), the quantum battery is initially prepared such that its total ergotropy originates entirely from coherence. This choice avoids repetition, as the aim of this section is to  analyze numerically the bound in Eq.~\ref{COH_COR_Bound}.
	
	In Fig.~\ref{EQ33}, we present a parametric plot of the coherence-based ergotropy of the quantum battery, namely $\mathcal{E}B^C(t)$, as a function of the difference between the coherence energy and the correlation energy, $\mathcal{W}{\mathrm{coh}}(t) - \mathcal{W}_{\mathrm{cor}}(t)$, in order to numerically verify Eq.~\ref{COH_COR_Bound}.
	
    For all scenarios ($I_c$, $II_c$, $III_c$), the initial values are $\mathcal{E}_B^C(0)=0.5$ and $\mathcal{W}_{\mathrm{coh}}(0) - \mathcal{W}_{\mathrm{cor}}(0)=1$. Over time, the decay in ergotropy is accompanied by a decrease in $\mathcal{W}_{\mathrm{coh}}(t) - \mathcal{W}_{\mathrm{cor}}(t)$. When $\mathcal{E}_B^C(t)=0$, this corresponds to $\mathcal{W}_{\mathrm{coh}}(t) - \mathcal{W}_{\mathrm{cor}}(t)=0$.
	
	Linking these results to the bound in Eq.~\ref{COHERENCE BOUND}, one can obtain $\mathcal{W}(\hat{\rho}_B(t))=\mathcal{W}_{\mathrm{coh}}(t) - \mathcal{W}_{\mathrm{cor}}(t)$, where the condition $W(\hat{\rho}_B(t))=0$ implies that the coherence-based ergotropy is bounded as: 
		\begin{align}
			0 \leq \mathcal{E}_B^{C} \leq W(\hat{\rho}_B) := K_B T\, \mathcal{C}(\hat{\rho}_B)=0.
	\end{align}
	
	Numerically, we find that $\mathcal{E}_B^C(t)=0$ coincides with $\mathcal{W}_{\mathrm{coh}}(t) - \mathcal{W}_{\mathrm{cor}}(t)=0$, which confirms that Eq.~\ref{COH_COR_Bound} satisfies explicitly the bound in Eq.~\ref{COHERENCE BOUND}.
	
	\begin{figure*}[htp!]
		
		\subfloat[  \label{Ex_1_33}]{%
			\includegraphics[width=0.58\columnwidth]{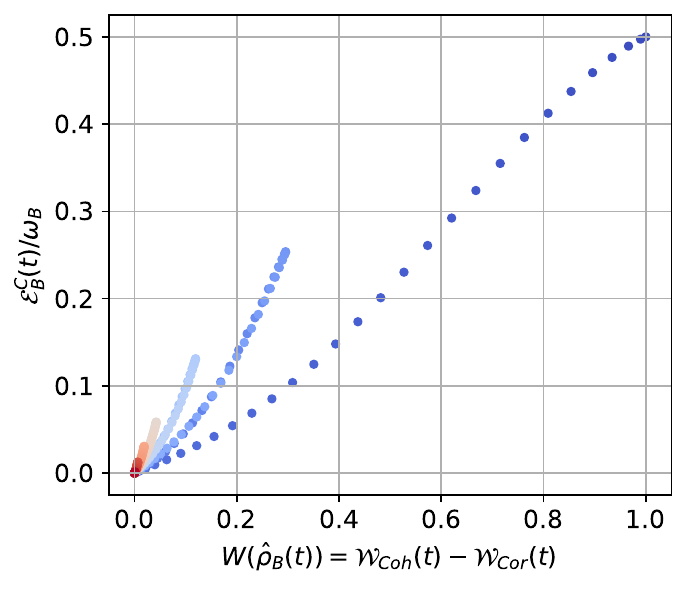}%
		}\hfill
		\subfloat[\label{EX_2_33}]{%
			\includegraphics[width=0.59\columnwidth]{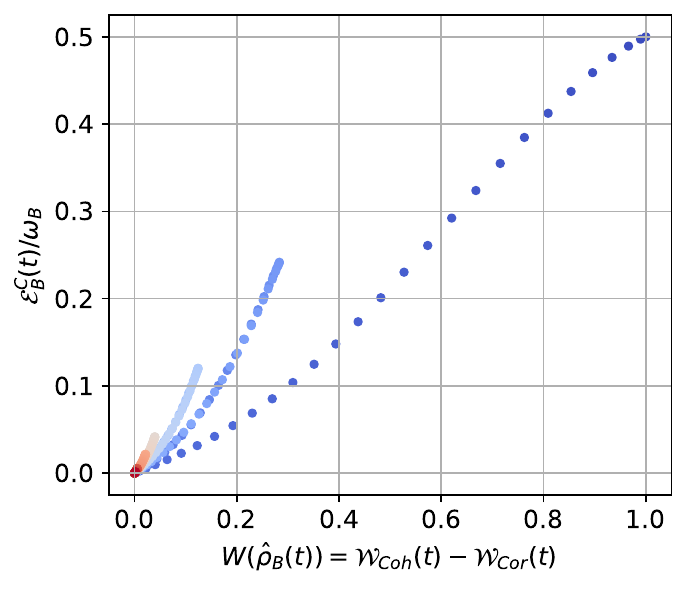}%
		}\hfill
		\subfloat[\label{EX_3_33}]{%
			\includegraphics[width=0.69\columnwidth]{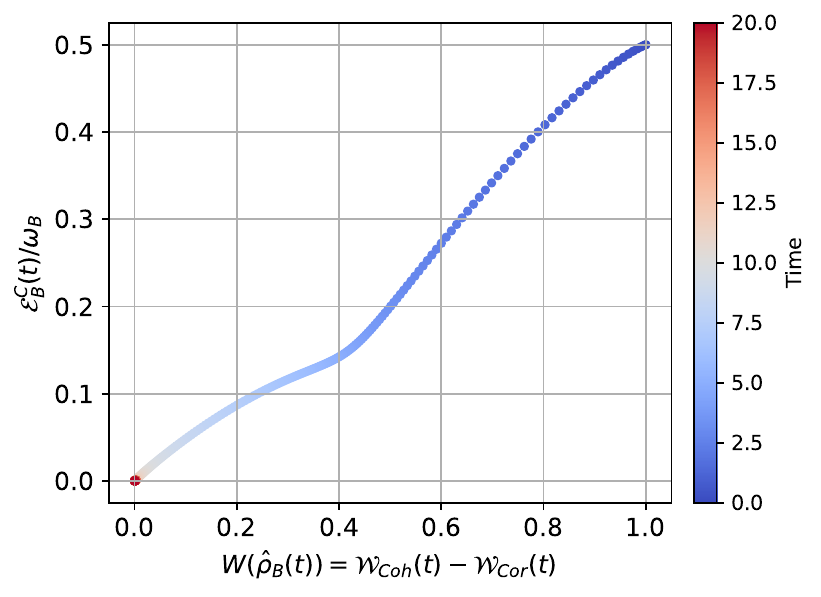}%
		}

		\caption{Parametric plots of ergotropy of coherence of quantum battery $\mathcal{E}_B^C(t)$ as a function of the difference between coherence energy and correlation energy, $\mathcal{W}_{\mathrm{coh}}(t) - \mathcal{W}_{\mathrm{cor}}(t)$, in panels (a)–(c) using scenarios ($I_c$, $II_c$, $III_c$). Besides, we set $g = 0.09$ and $k = 0.03$ (in units of $\omega_{S_2}$).}\label{EQ33}	
	\end{figure*}	
	
	\subsection{Numerical analysis of Eq.~\ref{RGO_COH_COR_Bounds}}\label{Numerical34}	
	Equation~\ref{RGO_COH_COR_Bounds} describes the upper and lower bounds of the total ergotropy of the quantum battery over time, which can be written as follows: 
		\begin{align}
			\mathcal{E}_B^{\mathrm{Lower}}(t) &\leq \mathcal{E}_B(t) \leq \mathcal{E}_B^{\mathrm{Upper}}(t), \nonumber\\
			\mathcal{E}_B^{\mathrm{Upper}}(t) &= \mathcal{E}_B^P(t) + \mathcal{W}_{\mathrm{coh}}(t) - \mathcal{W}_{\mathrm{cor}}(t), \nonumber\\
			\mathcal{E}_B^{\mathrm{Lower}}(t) &= \mathcal{E}_B^P(t).
	\end{align}
	
	this section aims to analyze numerically the upper and lower bounds of ergotropy of quantum battery over time, where we consider the initial states corresponding to the scenarios ($I_c$, $II_c$, $III_c$).
	
	In Fig.~\ref{EQ34}, we investigate numerically these bounds. Initially, the lower bound vanishes, i.e., $\mathcal{E}_B^{\mathrm{Lower}}(0)=\mathcal{E}_B^P(0)=0$. While, the upper bound takes a finite value, leading to: 
		\begin{align}
			0 \leq \mathcal{E}_B(t) \leq \mathcal{E}_B^{\mathrm{Upper}}(t).
	\end{align}
	
	Physically, this indicates that the system initially stores a finite amount of extractable work in terms of quantum coherence, such that $\mathcal{W}_{\mathrm{coh}}(0) > \mathcal{W}_{\mathrm{cor}}(0)$. Therefore, the ergotropy is initially dominated by coherence contributions, while the population contribution vanishes.
	
	Obviously, the increase of the lower bound reflects the gradual buildup of population inversion in the quantum battery. This indicates that the charging process shifts progressively from a coherence-dominated regime to a population-driven regime, while coherence is partially degraded due to reservoir-induced decoherence.
	
	For large values of times ($t \geq 10$), we obtain
		\begin{align}
			\mathcal{E}_B^{\mathrm{Lower}}(t) = \mathcal{E}_B(t) = \mathcal{E}_B^{\mathrm{Upper}}(t),
		\end{align}
		which shows that the total ergotropy originates entirely from population, i.e., $\mathcal{E}_B(t) = \mathcal{E}_B^P(t)$. This demonstrates that ergotropy bounds capture both its quantitative evolution and transition from a coherence-dominated to a population-dominated charging regime.

	\begin{figure*}[htp!]
		
		\subfloat[  \label{EX_1_NUM}]{%
			\includegraphics[width=0.58\columnwidth]{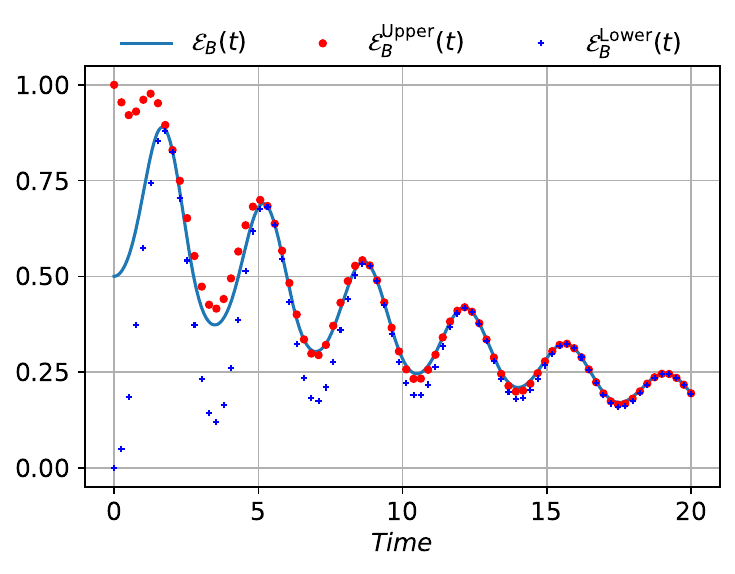}%
		}\hfill
		\subfloat[\label{EX_2_NUM}]{%
			\includegraphics[width=0.59\columnwidth]{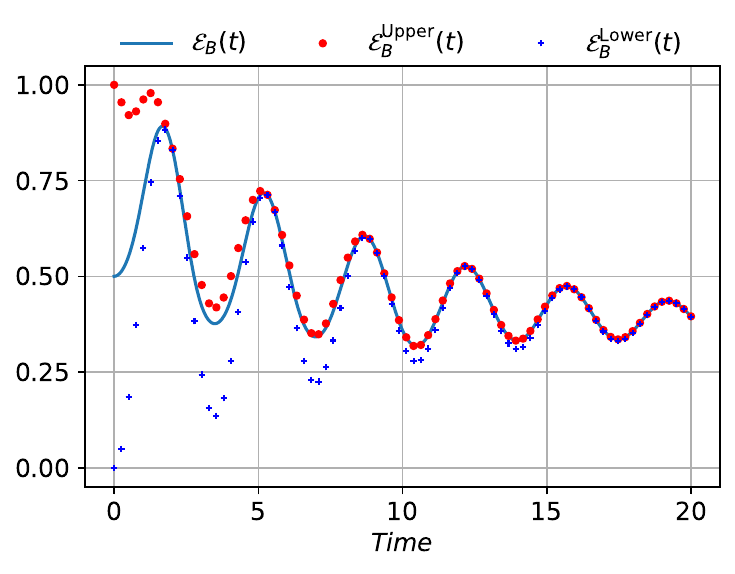}%
		}\hfill
		\subfloat[\label{EX_3_NUM}]{%
			\includegraphics[width=0.59\columnwidth]{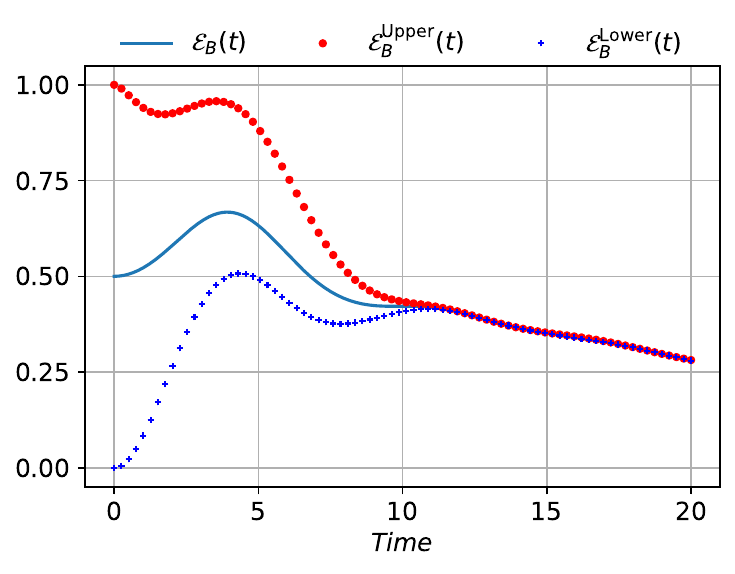}%
		}
		
		\caption{Dynamics of $\mathcal{E}_{B}(t)/\omega_{B}$, $\mathcal{E}_B^{\mathrm{Upper}}(t)/\omega_{B}$, and $\mathcal{E}_B^{\mathrm{Lower}}(t)/\omega_{B}$ in panels (a)–(c). 
				The panels correspond to the scenarios ($I_c$, $II_c$, $III_c$). 
				The parameters are set to $g = 0.09$ and $k = 0.03$ (in units of $\omega_{S_2}$). 
				This figure numerically analyzes the validity of Eq.~\ref{RGO_COH_COR_Bounds}.}\label{EQ34}	
	\end{figure*}	
	
	\section{Effect of $g$ and $k$ on the ergotropy of quantum battery}
\label{effectofgandk}
\begin{figure*}[htp!]
	\subfloat[  \label{RGO1_Density_classic}]{%
		\includegraphics[width=0.518\columnwidth]{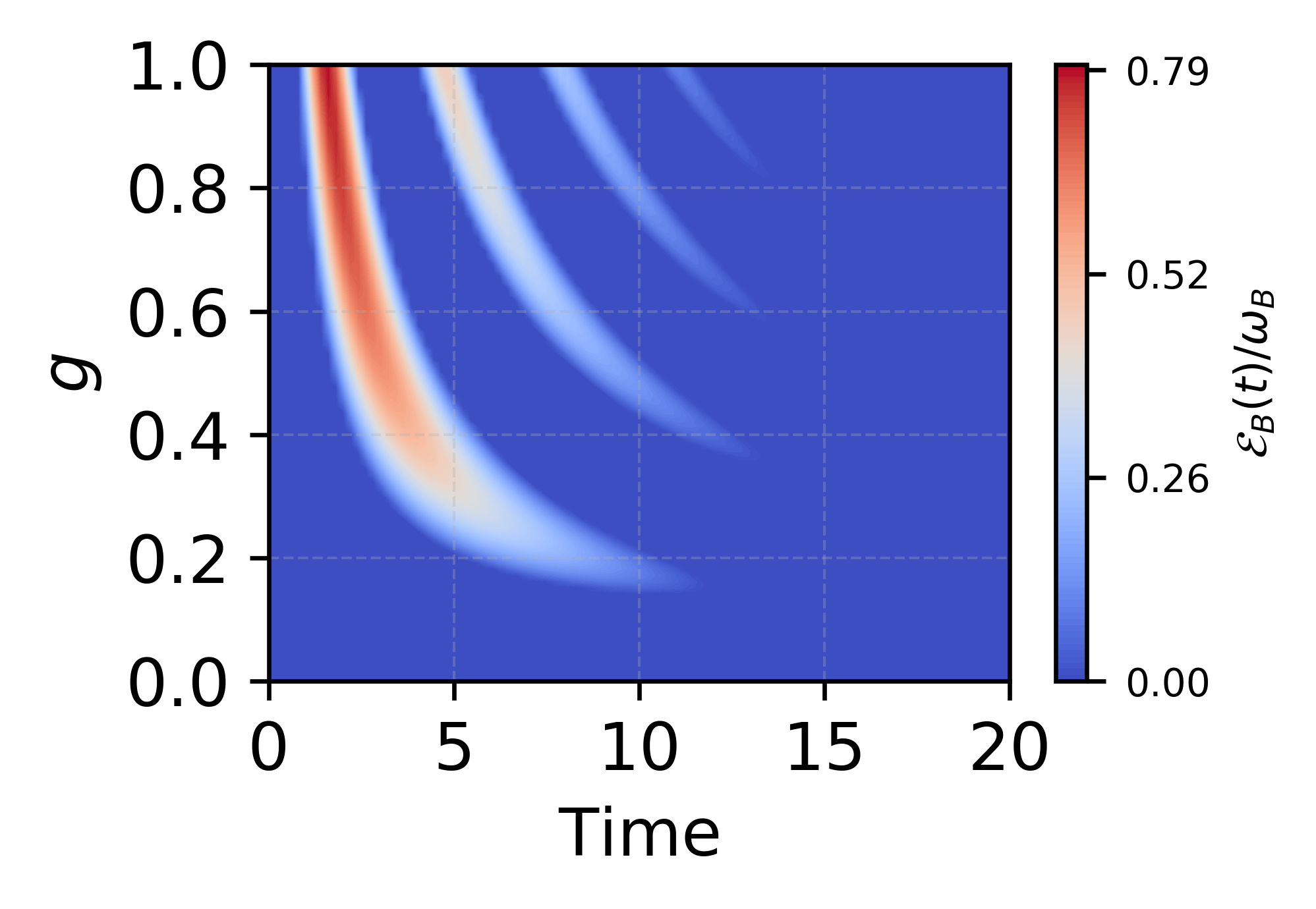}%
	}\hfill
	\subfloat[\label{RGO2_Density_classic}]{%
		\includegraphics[width=0.518\columnwidth]{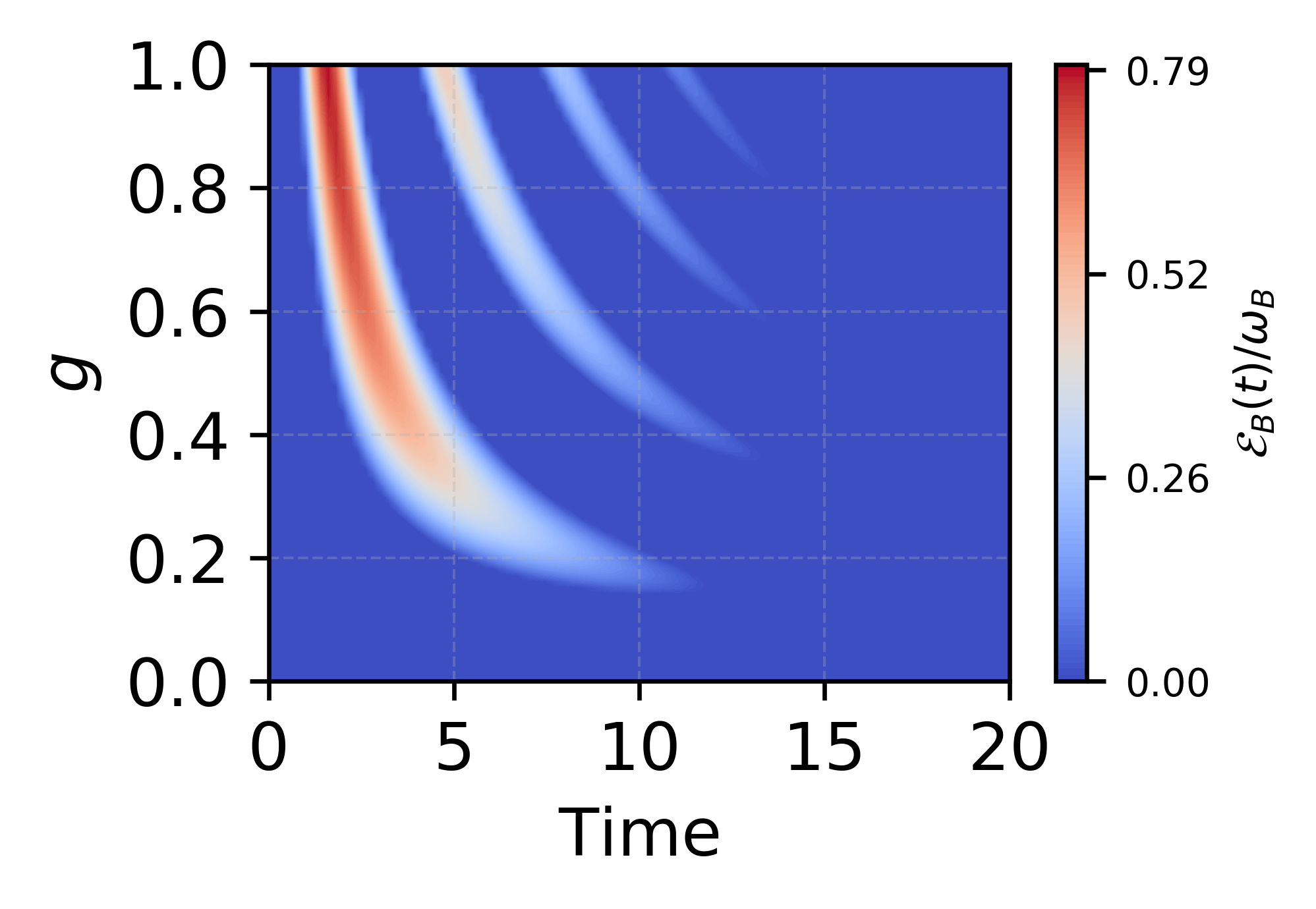}%
	}\hfill
	\subfloat[\label{RGO3_Density_classic_g}]{%
		\includegraphics[width=0.518\columnwidth]{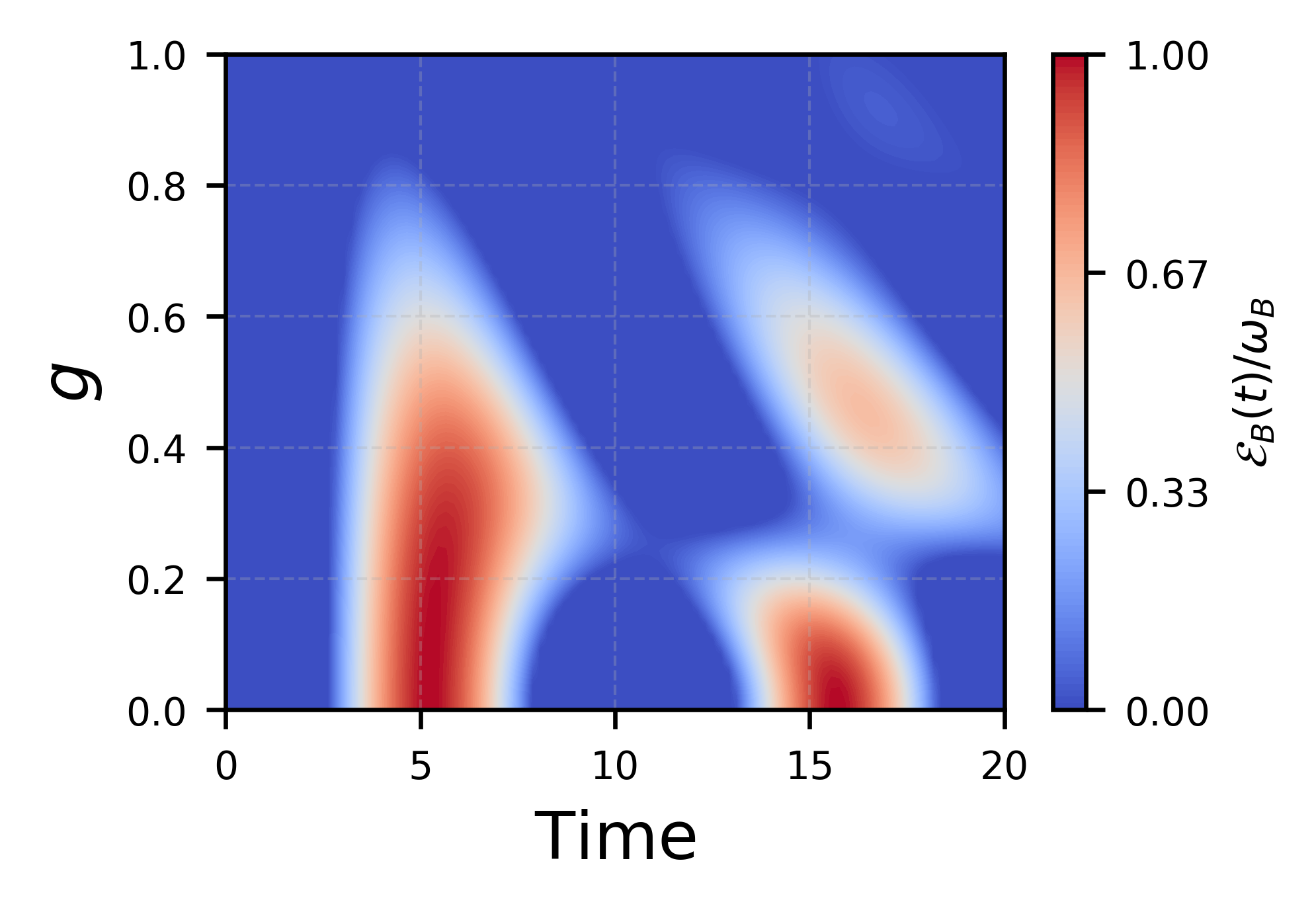}%
	}
	\hfill
	\subfloat[\label{RGO3_Density_classic_k}]{%
		\includegraphics[width=0.518\columnwidth]{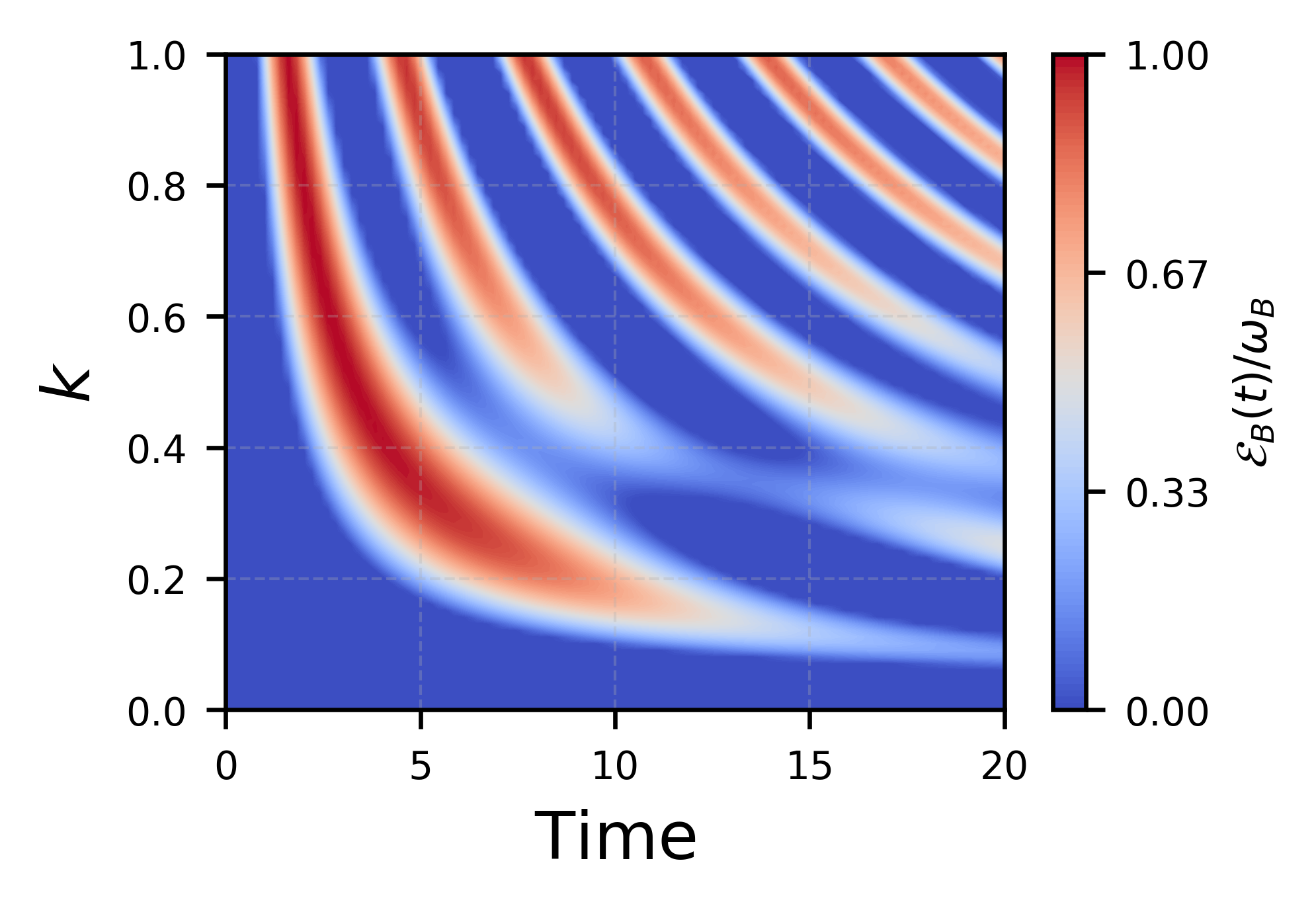}%
	}\hfill
	\subfloat[\label{RGO1_Density_quantum}]{%
		\includegraphics[width=0.518\columnwidth]{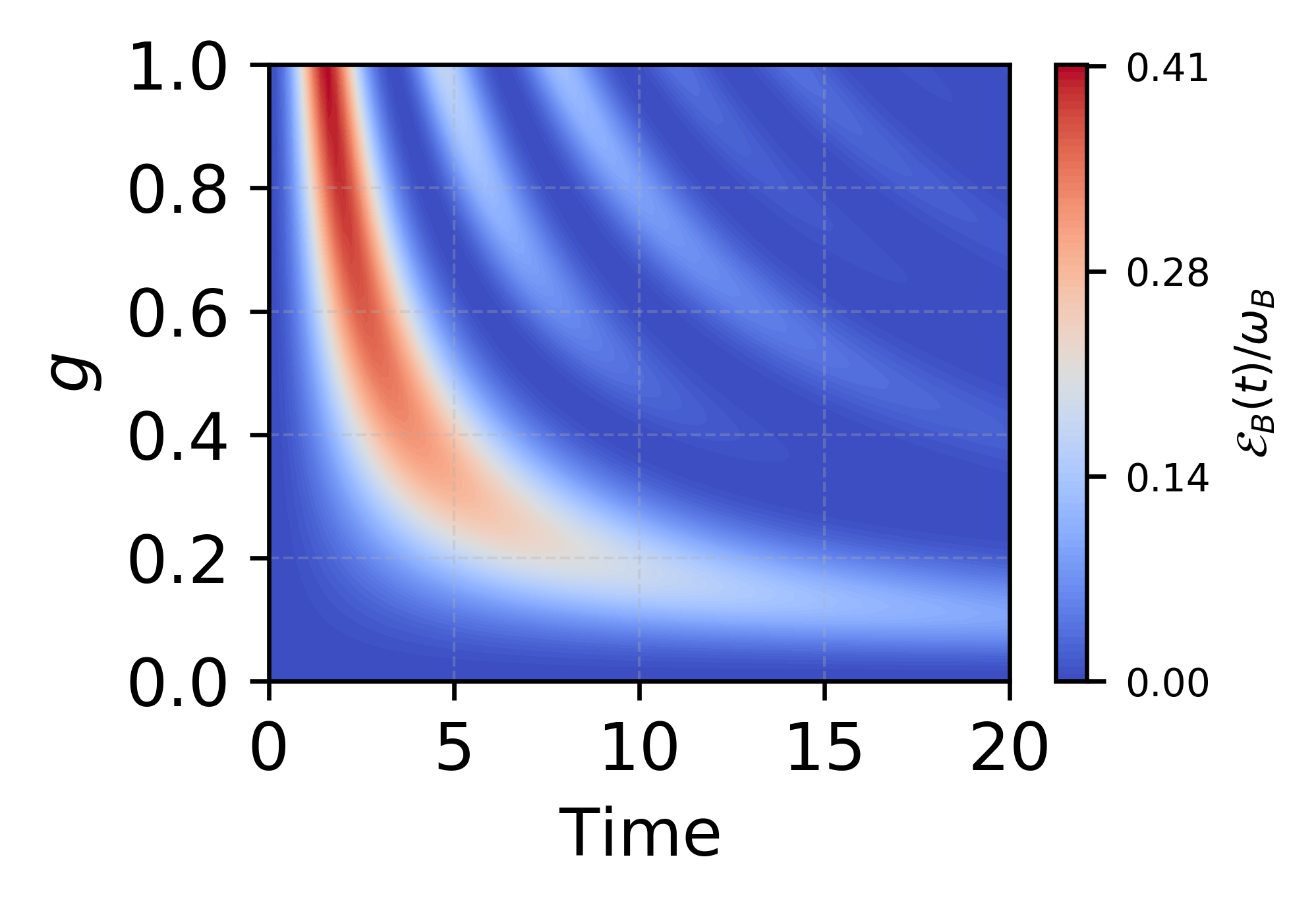}%
	}\hfill
	\subfloat[\label{RGO2_Density_quantum}]{%
		\includegraphics[width=0.518\columnwidth]{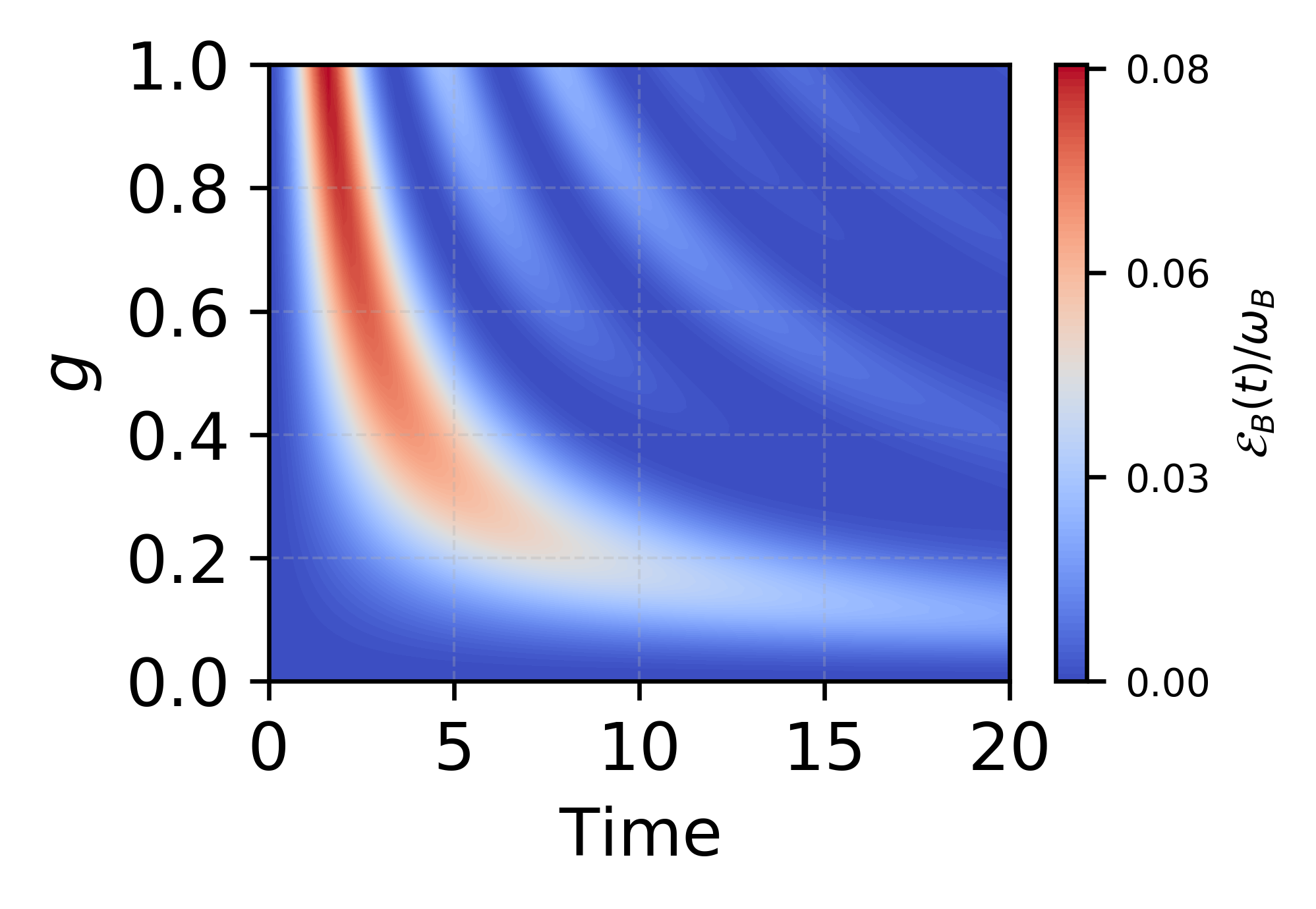}%
	}
	\hfill
	\subfloat[\label{RGO3_Density_quantum_g}]{%
		\includegraphics[width=0.518\columnwidth]{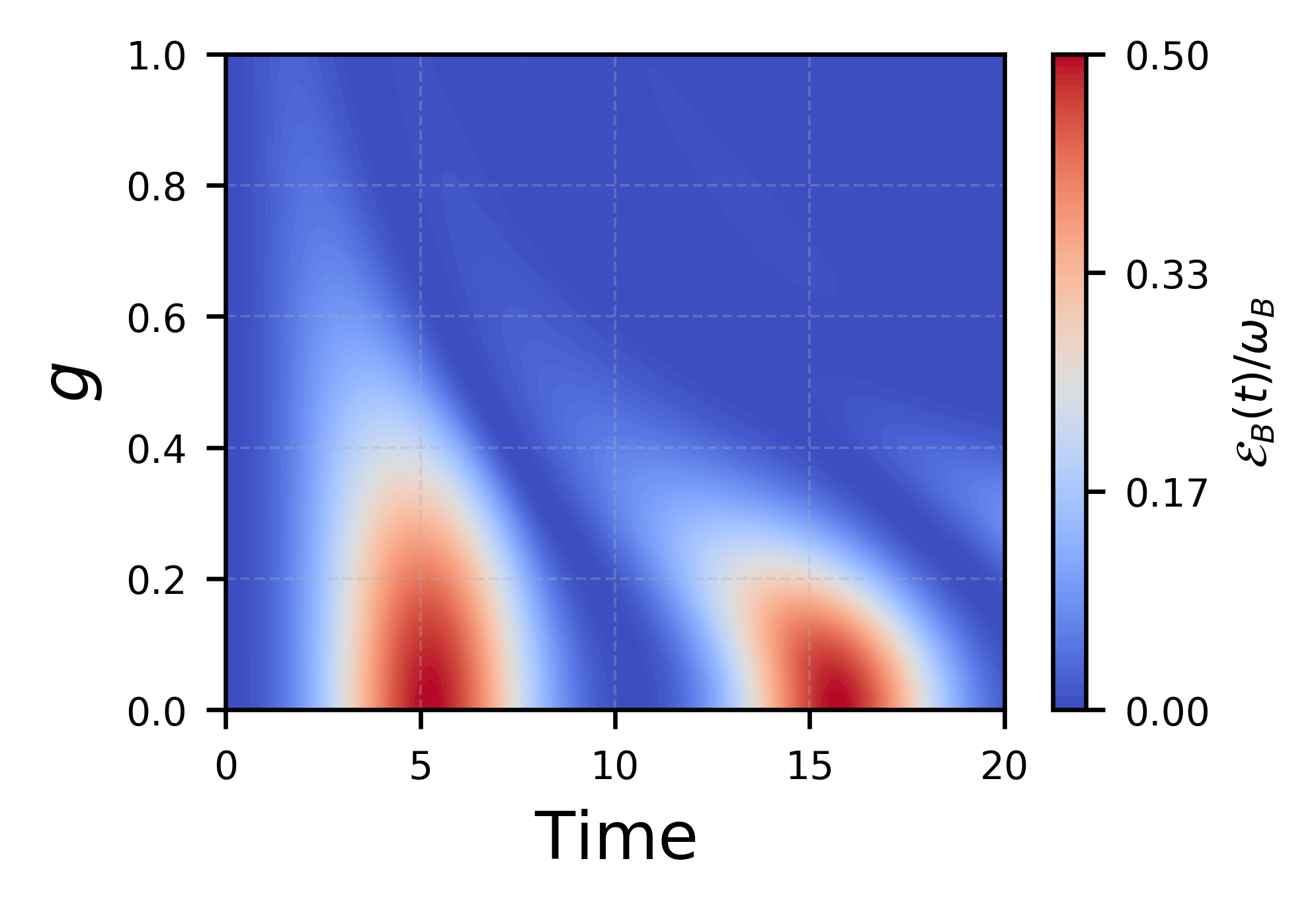}%
	}\hfill
	\subfloat[\label{RGO3_Density_quantum_k}]{%
		\includegraphics[width=0.518\columnwidth]{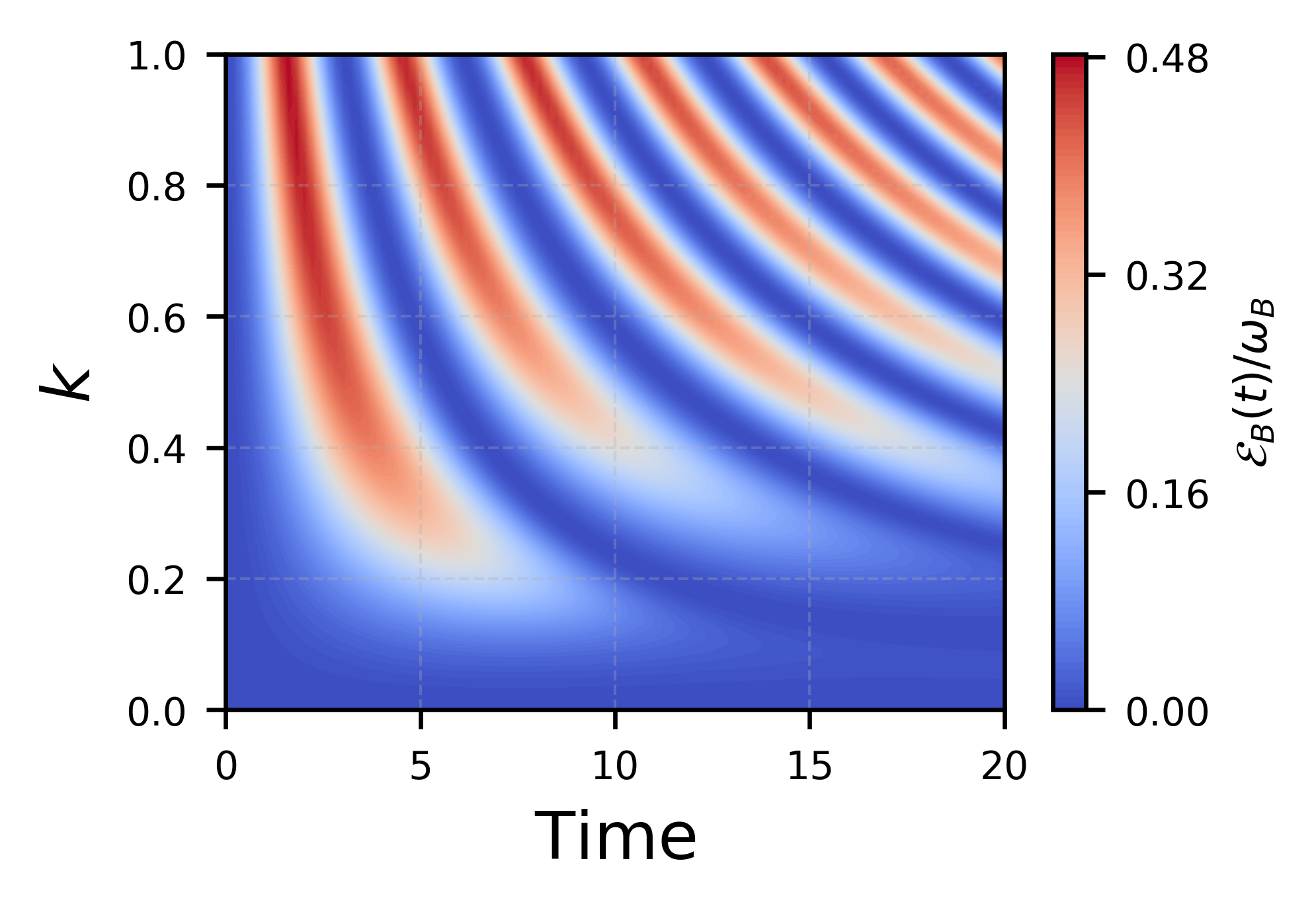}%
	}
	\caption{Density plots of quantum battery ergotropy $\mathcal{E}_B(t)$ as a function of time and coupling strengths $g$ and $k$. Panels (a,b,c,d) correspond to the scenarios $I_a$, $II_a$, $III_a$ (we set $k=0.03\omega_{S_2}$), $III_a$ (we set $g=0.03\omega_{S_2}$). Panels (e,f,g,h) correspond to the scenarios $I_b$, $II_b$, $III_b$ (we set $k=0.03\omega_{S_2}$), $III_b$ (we set $g=0.03\omega_{S_2}$). The physical aim is to show the effect of variation in the coupling strengths on the time evolution of the ergotropy.}	\label{ERGO_VS_g_k}	
\end{figure*}
	The interaction Hamiltonian $\hat{H}_{\mathrm{Int}_\alpha}$ plays a crucial role in the generation of correlations and the transfer of coherence between the subsystems, as it governs the system dynamics through the dissipative term of the master equation in Eq.~\ref{LL}. 
	
	Figure~\ref{Coherence_energy} provides a direct quantitative illustration of how the interaction strengths $g$ and $k$ control the charging performance of the quantum battery through their impact on correlation and coherence dynamics. As shown in Fig.~\ref{Coherence_energy}, the energy stored in coherence increases when $g$ increases for scenarios $I_b$ and $II_b$. In contrast, an opposite behavior is observed in scenario $III_b$, where the coherence energy decreases as $g$ increases. This behavior directly affects the total ergotropy of the quantum battery in agreement with the bound given in Eq.~\ref{RGO_COH_COR_Bounds}.
	
	In Fig.~\ref{ERGO_VS_g_k}, we further analyze the effect of the couplings $g$ and $k$ on the total ergotropy of the quantum battery. For scenarios $I_a$, $II_a$, $I_b$ and $II_b$, the increase in ergotropy is straightforwardly related to the increase in coupling strength between $S_{12}$, charger and battery. This behavior is consistent with the ergotropy bounds provided in Eq.~\ref{RGO_COH_COR_Bounds}.
	
	On the other hand, for scenario $III_a$ ($k = 0.03\,\omega_{S_2}$) and scenario $III_b$ ($g = 0.03\,\omega_{S_2}$), a different behavior is observed. When $k$ is fixed and $g$ is varied, the ergotropy decreases with increasing $g$. Conversely, when $g$ is fixed and $k$ is varied, the ergotropy increases with increasing $k$. Physically, this behavior arises from the interplay between the collective interaction involving $S_{12}$ and the charger, as well as the local interaction between the charger and the battery in scenario III. This structure provides greater controllability of the charging process compared to the other scenarios. Finally, note that the case of $g = 0$ corresponds to an idealized situation in which the charger--battery system is effectively isolated from the structured reservoir, where the decoherence effects induced by the reservoirs vanish.
	
	Focusing on the maximum ergotropy, we observe that scenarios $I_a$ and $II_a$ exhibit identical values, reaching $\mathcal{E}_B^{\max} \approx 0.79$ for $g \approx 0.1\,\omega_{S_2}$. In contrast, scenario $III_a$ achieves higher values, with $\mathcal{E}_B^{\max} \approx 1$ for $g = 0.03\,\omega_{S_2}$. While, by increasing $k$, a more efficient charging mechanism is observed.
	
	For the coherent initial state ($S_b$), scenario $II_b$ shows the lowest performance with $\mathcal{E}_B^{\max} \approx 0.08$, while scenario $I_b$ reaches moderate values. The highest ergotropy is obtained in scenario $III_b$, with $\mathcal{E}_B^{\max} \approx 0.48$ for $g = 0.03\,\omega_{S_2}$ and large $k$. Overall, these results demonstrate that increasing the charger--battery coupling $k$ enhances the maximum extractable work. Whereas, increasing $g$ in scenario III tends to reduce it, highlighting the distinct roles of $g$ and $k$ on the charging process.
	
	Notably, in scenarios $I_a$ and $II_a$, the interaction Hamiltonians describe resonant energy-exchange processes that effectively reduce to similar Rabi-type oscillations between collective states. Although, the microscopic transitions differ, the normalization of the ergotropy ($\mathcal{E}_B/\omega_B$) removes the energy scale, leading to nearly identical dynamics as observed in Figs.~\ref{RGO1_Density_classic} and \ref{RGO2_Density_classic}.
	
A similar argument can be applied to Figs.~\ref{RGO1_Density_quantum} and \ref{RGO2_Density_quantum}, where small effective coupling results in slower coherence transfer and reduced amplitudes. After normalization, the curves mainly differ by a scaling factor while preserving the same qualitative behavior.
	
	Overall, this similarity indicates that different interaction scenarios, despite involving distinct multipartite processes, give rise to comparable effective energy-exchange dynamics when expressed in normalized units.

\section{Decomposition of the Battery Energy Current}
\label{app:current_decomposition}

In this Appendix, we derive the decomposition of the energy current stored in the quantum battery into two contributions: a factorized local contribution and a correlation-induced contribution. This decomposition provides a quantitative way to identify how the correlations generated during the dynamics contribute to the energy stored in the battery.

We consider the total system $S=S_1\otimes S_2\otimes C\otimes B$, whose density matrix is denoted by $\hat{\rho}_{S}(t)$. The dynamics of $\hat{\rho}_{S}(t)$ is governed by the master equation introduced in Eq.~\eqref{MASTER_EQUATION}. In particular, the superoperator $\mathcal{L}_{S_{12}}\left[\hat{\rho}_{S}(t)\right]$ already defined in Eq.~\eqref{LL} contains both the coherent interaction between the subsystems and the dissipative contributions induced by the reservoirs. Here, the dissipation consists of two parts, where $-i\left[H_{Int_{\alpha}}, \hat{\rho}_S(t)\right]$ denotes the interaction corresponding to the charging scenario $\alpha=\{I,II,III\}$. Whereas, the second contribution, namely $\mathcal{D}^{[T_{R_m}]}$ describes the local dissipative action of the thermal reservoir $R_m$ on the structured-reservoir qubit $S_m$.

The battery energy is defined as
	\begin{align}
		E_B(t)
		&=
		\mathrm{Tr}_{B}
		\left\{
		\hat{\rho}_{B}(t)\hat{H}_{B}
		\right\}
		\nonumber\\
		&=
		\mathrm{Tr}_{S}
		\left\{
		\hat{\rho}_{S}(t)\hat{H}_{B}
		\right\}.
		\label{eq:app_battery_energy_current}
	\end{align}
	Therefore, the energy current stored in the battery is given as:
	\begin{align}
		J_B(t)
		&=
		\frac{d}{dt}E_B(t)
		\nonumber\\
		&=
		\mathrm{Tr}_{S}
		\left\{
		\frac{d}{dt}\hat{\rho}_{S}(t)\hat{H}_{B}
		\right\}.
		\label{eq:app_battery_current_def}
	\end{align}
Now, by substituting Eqs.~\eqref{MASTER_EQUATION} and ~\eqref{LL} into Eq.~\eqref{eq:app_battery_current_def}, one can obtain
	\begin{align}
		J_B(t)
		=&
		-i\,
		\mathrm{Tr}_{S}
		\left\{
		\left[
		\sum_{m=1}^{2}\hat{H}_{S_m}
		+
		\sum_{n=\{C,B\}}\hat{H}_{n},
		\hat{\rho}_{S}(t)
		\right]
		\hat{H}_{B}
		\right\}
		\nonumber\\
		&
		-i\,
		\mathrm{Tr}_{S}
		\left\{
		\left[
		\hat{H}_{\mathrm{Int}_{\alpha}},
		\hat{\rho}_{S}(t)
		\right]
		\hat{H}_{B}
		\right\}
		\nonumber\\
		&
		+
		\sum_{m=1}^{2}
		\mathrm{Tr}_{S}
		\left\{
		\mathcal{D}^{[T_{R_m}]}
		\left[
		\hat{\rho}_{S}(t)
		\right]
		\hat{H}_{B}
		\right\}.
		\label{eq:app_current_full}
	\end{align}
	The contribution of the free Hamiltonian vanishes because
	\begin{align}
		\mathrm{Tr}_{S}
		\left\{
		\left[
		\sum_{m=1}^{2}\hat{H}_{S_m}
		+
		\sum_{n=\{C,B\}}\hat{H}_{n},
		\hat{\rho}_{S}(t)
		\right]
		\hat{H}_{B}
		\right\}
		&=0,\nonumber\\
		\mathrm{Tr}_{S}
		\left\{
		\hat{\rho}_{S}(t)
		\left[
		\hat{H}_{B},
		\sum_{m=1}^{2}\hat{H}_{S_m}
		+
		\sum_{n=\{C,B\}}\hat{H}_{n}
		\right]
		\right\}
		&=
		0.	\nonumber
		\label{eq:app_free_current_zero}
	\end{align}
	In the last equality we used the fact that the free Hamiltonians act locally on different subsystems. Therefore
	\begin{equation}
		\left[
		\hat{H}_{B},
		\sum_{m=1}^{2}\hat{H}_{S_m}
		+
		\sum_{n=\{C,B\}}\hat{H}_{n}
		\right]
		=0.
	\end{equation}
	Moreover, the dissipators $\mathcal{D}^{[T_{R_m}]}$ act locally on the structured-reservoir qubits $S_m$ and not directly on the battery. Thus, their direct contribution to the battery energy current vanishes:
	\begin{align}
		\mathrm{Tr}_{S}
		\left\{
		\mathcal{D}^{[T_{R_m}]}
		\left[
		\hat{\rho}_{S}(t)
		\right]
		\hat{H}_{B}
		\right\}
		=
		0.
		\label{eq:app_dissipator_current_zero}
	\end{align}
	Hence, the reservoirs affect battery current indirectly through the time dependence of the total state $\hat{\rho}_{S}(t)$. While, the direct energy current entering the battery is generated by the interaction Hamiltonian $\hat{H}_{\mathrm{Int}_{\alpha}}$. Consequently,
	\begin{align}
		J_B(t)
		=
		-i\,
		\mathrm{Tr}_{S}
		\left\{
		\left[
		\hat{H}_{\mathrm{Int}_{\alpha}},
		\hat{\rho}_{S}(t)
		\right]
		\hat{H}_{B}
		\right\}.
		\label{eq:app_current_interaction_only}
	\end{align}

To distinguish the part of the current generated by correlations, we decompose the total state as
	\begin{align}
		\hat{\rho}_{S}(t)
		=
		\hat{\rho}_{S_{12}}(t)
		\otimes
		\hat{\rho}_{C}(t)
		\otimes
		\hat{\rho}_{B}(t)
		+
		\hat{\rho}_{\mathrm{cor}}(t),
		\label{eq:app_state_decomposition_current}
	\end{align}
	where the correlation operator is defined the following formula: 
	\begin{align}
		\hat{\rho}_{\mathrm{cor}}(t)
		=
		\hat{\rho}_{S}(t)
		-
		\hat{\rho}_{S_{12}}(t)
		\otimes
		\hat{\rho}_{C}(t)
		\otimes
		\hat{\rho}_{B}(t).
		\label{eq:app_correlation_operator_current}
	\end{align}
	Here, $\hat{\rho}_{S_{12}}(t)$ denotes the reduced state of the structured reservoir qubits $S_1$ and $S_2$. Therefore, $\hat{\rho}_{\mathrm{cor}}(t)$ contains the correlations established between the block $S_{12}$, the charger $C$ and the battery $B$ during the charging dynamics.

Replacing Eq.~\eqref{eq:app_state_decomposition_current} into Eq.~\eqref{eq:app_current_interaction_only}, one can obtain
	\begin{align}
		J_B(t)
		=&
		-i\,
		\mathrm{Tr}_{S}
		\left\{
		\left[
		\hat{H}_{\mathrm{Int}_{\alpha}},
		\hat{\rho}_{S_{12}}(t)
		\otimes
		\hat{\rho}_{C}(t)
		\otimes
		\hat{\rho}_{B}(t)
		\right]
		\hat{H}_{B}
		\right\}
		\nonumber\\
		&
		-i\,
		\mathrm{Tr}_{S}
		\left\{
		\left[
		\hat{H}_{\mathrm{Int}_{\alpha}},
		\hat{\rho}_{\mathrm{cor}}(t)
		\right]
		\hat{H}_{B}
		\right\}.
		\label{eq:app_current_split}
	\end{align}
	Accordingly, we define the factorized local contribution to the battery current as
	\begin{align}
		J_B^{\mathrm{loc}}(t)
		=
		-i\,
		\mathrm{Tr}_{S}
		\left\{
		\left[
		\hat{H}_{\mathrm{Int}_{\alpha}},
		\hat{\rho}_{S_{12}}(t)
		\otimes
		\hat{\rho}_{C}(t)
		\otimes
		\hat{\rho}_{B}(t)
		\right]
		\hat{H}_{B}
		\right\}.
		\label{eq:app_current_local}
	\end{align}
While, the correlation-induced contribution reads as:
	\begin{align}
		J_B^{\mathrm{cor}}(t)
		=
		-i\,
		\mathrm{Tr}_{S}
		\left\{
		\left[
		\hat{H}_{\mathrm{Int}_{\alpha}},
		\hat{\rho}_{\mathrm{cor}}(t)
		\right]
		\hat{H}_{B}
		\right\}.
		\label{eq:app_current_correlation}
	\end{align}
	The total energy current stored in the battery is therefore decomposed as
	\begin{align}
		J_B(t)
		=
		J_B^{\mathrm{loc}}(t)
		+
		J_B^{\mathrm{cor}}(t).
		\label{eq:app_total_current_decomposition}
	\end{align}

The total variation of the stored energy in the battery is
	\begin{align}
		\Delta E_B(t)
		&=
		E_B(t)-E_B(0)
		\nonumber\\
		&=
		\int_{0}^{t}J_B(t')\,dt'.
		\label{eq:app_delta_E_total}
	\end{align}
	Using Eq.~\eqref{eq:app_total_current_decomposition}, this variation can be written as
	\begin{align}
		\Delta E_B(t)
		=
		\Delta E_B^{\mathrm{loc}}(t)
		+
		\Delta E_B^{\mathrm{cor}}(t),
		\label{eq:app_delta_E_decomposition}
	\end{align}
	where
	\begin{align}
		\Delta E_B^{\mathrm{loc}}(t)
		=
		\int_{0}^{t}
		J_B^{\mathrm{loc}}(t')\,dt',
		\label{eq:app_delta_E_local}
	\end{align}
	and
	\begin{align}
		\Delta E_B^{\mathrm{cor}}(t)
		=
		\int_{0}^{t}
		J_B^{\mathrm{cor}}(t')\,dt'.
		\label{eq:app_delta_E_correlation}
	\end{align}
	This decomposition shows that the energy stored in the battery can be separated into a contribution generated by the factorized local dynamics of the reduced states and another contribution generated by the correlations exchanged between structured reservoir, charger, and battery. In this regards, $J_B^{\mathrm{cor}}(t)$ and $\Delta E_B^{\mathrm{cor}}(t)$ provide a direct quantitative measure of the role of dynamical correlations in the autonomous charging process.


\begin{thebibliography}{9}
		
		\bibitem{intro1}
		Max F Riedel et al, "The European quantum technologies flagship programme
		" , Quantum Sci. Technol. \textbf{2} 030501 (2017).
		\bibitem{intro2}	
		Antonio Acín et al,"The quantum technologies roadmap: a European community view",  New J. Phys. \textbf{20} 080201 (2018).	
		\bibitem{intro3}
		J.~Q.~Quach, G.~Cerullo, and T.~Virgili,
		``Quantum batteries: The future of energy storage?,'' 
		\textit{Joule} \textbf{7}, 2195--2200 (2023).
		\bibitem{intro4}	
		A.~Demir, E.~Yildiz, and C.~Kaya,
		``Application of quantum computing in the design of new materials for batteries,''
		\textit{J. Tecnol. Quantica} \textbf{1}, 288--300 (2024).
		\bibitem{intro5}
		D.~Farina, G.~M.~Andolina, A.~Mari, M.~Polini, and V.~Giovannetti,
		``Charger-mediated energy transfer for quantum batteries: An open-system approach,''
		\textit{Phys. Rev. B} \textbf{99}, 035421 (2019).
		\bibitem{intro6}
		G.~M.~Andolina, D.~Farina, A.~Mari, V.~Pellegrini, V.~Giovannetti, and M.~Polini,
		``Charger-mediated energy transfer in exactly solvable models for quantum batteries,''
		\textit{Phys. Rev. B} \textbf{98}, 205423 (2018).
		\bibitem{intro6a}
     C.-J.~Wang and F.-Q.~Dou,
	``Ergotropy in Quantum Batteries,''
	\textit{Chin. Phys. Lett.} \textbf{43}, 060602 (2026).
		\bibitem{intro7}
		S.~Elghaayda, A.~Ali, S.~Al-Kuwari, A.~Czerwinski, M.~Mansour, and S.~Haddadi,
		``Performance of a superconducting quantum battery,'' 		\textit{Adv. Quantum Technol.} \textbf{2025}, 2400651.
		
		\bibitem{intro7_a}
	F.-M.~Yang and F.-Q.~Dou,
	``Resonator-qutrit quantum battery,''
	\textit{Phys. Rev. A} \textbf{109}, 062432 (2024).
		
		\bibitem{intro7_b}
F.-Q.~Dou and F.-M.~Yang,
	``Superconducting transmon qubit-resonator quantum battery,''
	\textit{Phys. Rev. A} \textbf{107}, 023725 (2023).	

		\bibitem{intro8}
		Z.~G.~Lu, G.~Tian, X.~Y.~Lü, and C.~Shang,
		``Topological quantum batteries,''
		\textit{Phys. Rev. Lett.} \textbf{134}, 180401 (2025).
		\bibitem{intro9}
		B.~Ahmadi, P.~Mazurek, S.~Barzanjeh, and P.~Horodecki,
		``Super-optimal charging of quantum batteries via reservoir engineering,''
		\textit{Phys. Rev. Appl.} \textbf{23}, 024010 (2024).
		\bibitem{intro9_a}
	H.-L.~Shi, S.~Ding, Q.-K.~Wan, X.-H.~Wang, and W.-L.~Yang,
	``Entanglement, coherence, and extractable work in quantum batteries,''
	\textit{Phys. Rev. Lett.} \textbf{129}, 130602 (2022).
		
		
		\bibitem{intro10}
		F.~Cavaliere, G.~Gemme, G.~Benenti, D.~Ferraro, and M.~Sassetti,
		``Dynamical blockade of a reservoir for optimal performances of a quantum battery,''
		\textit{Commun. Phys.} \textbf{8}, 76 (2025).
		\bibitem{intro10_a}
		A.~Khoudiri, A.~Ullah, A.~El Allati, and Ö.~E.~Müstecaplıoğlu,
		``Collective-dissipation-induced dark and metastable-like states for enhanced quantum battery performance,''
		\textit{arXiv preprint arXiv:2608.09693} (2026).
		\bibitem{intro11}
		Y.~Yao and X.~Q.~Shao,
		``Reservoir-assisted quantum battery charging at finite temperatures,''
		\textit{Phys. Rev. A} \textbf{111}, 062616 (2025).
		
		\bibitem{intro12}
		B.~Ahmadi, P.~Mazurek, P.~Horodecki, and S.~Barzanjeh,
		``Nonreciprocal quantum batteries,''
		\textit{Phys. Rev. Lett.} \textbf{132}, 210402 (2024).
		\bibitem{intro13}
		K.~Xu, H.~J.~Zhu, G.~F.~Zhang, and W.~M.~Liu,
		``Enhancing the performance of an open quantum battery via environment engineering,''
		\textit{Phys. Rev. E} \textbf{104}, 064143 (2021).
		\bibitem{intro14}
		J.~L.~Li, H.~Z.~Shen, and X.~X.~Yi,
		``Quantum batteries in non-Markovian reservoirs,''
		\textit{Opt. Lett.} \textbf{47}, 5614--5617 (2022).
		\bibitem{intro14_a}
	W.-L.~Song, J.-L.~Wang, B.~Zhou, W.-L.~Yang, and J.-H.~An,
	``Self-discharging mitigated quantum battery,''
	\textit{Phys. Rev. Lett.} \textbf{135}, 020405 (2025).
		\bibitem{INTRO_rev_1}
		T.~M.~Mendonça, A.~M.~Souza, R.~J.~de~Assis, N.~G.~de~Almeida, R.~S.~Sarthour, I.~S.~Oliveira, and C.~J.~Villas-Boas,
			``Reservoir engineering for maximally efficient quantum engines,''
			\textit{Phys. Rev. Res.} \textbf{2}, 043419 (2020).
		
		\bibitem{INTRO_rev_2}
		J.~Yang, S.~Fang, C.~Zhao, C.~Shan, and B.~Xiong,
			``Quantum battery via reservoir engineering,''
			\textit{Phys. Rev. Appl.} \textbf{25}, 034013 (2026).
		
		\bibitem{INTRO_rev_3}
		F.~T.~Tabesh, F.~H.~Kamin, and S.~Salimi,
			``Environment-mediated charging process of quantum batteries,''
			\textit{Phys. Rev. A} \textbf{102}, 052223 (2020).
		\bibitem{INTRO_rev_4}
		M.~L.~Hu, T.~Gao, and H.~Fan,
	``Efficient wireless charging of a quantum battery,''
	\textit{Phys. Rev. A} \textbf{111}, 042216 (2025).
		\bibitem{INTRO_rev_5}
		M.-L.~Song, L.-J.~Li, X.-K.~Song, L.~Ye, and D.~Wang,
			``Environment-mediated entropic uncertainty in charging quantum batteries,''
			\textit{Phys. Rev. E} \textbf{106}, 054107 (2022).
		\bibitem{INTRO_rev_6}F.~Barra,
			``Dissipative charging of a quantum battery,''
			\textit{Phys. Rev. Lett.} \textbf{122}, 210601 (2019).
		\bibitem{MODEL8}
		M.~Alimuddin, T.~Guha, and P.~Parashar,
		``Structure of passive states and its implication in charging quantum batteries,''
		\textit{Phys. Rev. E} \textbf{102}, 022106 (2020).
		
		\bibitem{MODEL10}
	G.~Francica, F.~C.~Binder, G.~Guarnieri, M.~T.~Mitchison, J.~Goold, and F.~Plastina,
	``Quantum coherence and ergotropy,''
	\textit{Phys. Rev. Lett.} \textbf{125}, 180603 (2020).
		
		\bibitem{MODEL12}
		A.~Touil, B.~\c{C}akmak, and S.~Deffner,
		``Ergotropy from quantum and classical correlations,''
		\textit{J. Phys. A: Math. Theor.} \textbf{55}, 025301 (2021).
		\bibitem{MODEL13}
		T.~Biswas, M.~\L{}obejko, P.~Mazurek, K.~Ja\l{}owiecki, and M.~Horodecki,
		``Extraction of ergotropy: Free energy bound and application to open cycle engines,''
		\textit{Quantum} \textbf{6}, 841 (2022).
		\bibitem{MODEL9}
		R.~Castellano, D.~Farina, V.~Giovannetti, and A.~Acin,
		``Extended local ergotropy,''
		\textit{Phys. Rev. Lett.} \textbf{133}, 150402 (2024).
		
		\bibitem{MODEL1a}
		G.~G.~Damas, R.~J.~de~Assis, and N.~G.~de~Almeida,
		``Cooling with fermionic thermal reservoirs,''
		\textit{Physical Review E} \textbf{107}, 034128 (2023).
		\bibitem{MODEL2a}
		G.~Manzano,
		``Squeezed thermal reservoir as a generalized equilibrium reservoir,''
		\textit{Physical Review E} \textbf{98}, 042123 (2018).
		\bibitem{MODEL1} 	
		S.Lorenzo et al, "Composite quantum collision models", Phys. Rev. A, \textbf{96}, 032107 (2017).
		\bibitem{MODEL2}
		H.-P.Breuer, F.Petruccione. "The theory of open quantum systems". Oxford University Press(2002).
		\bibitem{MODEL3}
		A.Rivas Vargas, "Open quantum systems and quantum information dynamics", (Doctoral thesis, Universitaire Ulm) (2011).
		\bibitem{MODEL3a}
		C.~Fleming, N.~I.~Cummings, C.~Anastopoulos, and B.~L.~Hu,
		``The rotating-wave approximation: consistency and applicability from an open quantum system analysis,''
		\textit{Journal of Physics A: Mathematical and Theoretical} \textbf{43}, 405304 (2010).		
		\bibitem{MODEL4}
		S.~Ghosh, A.~Opala, M.~Matuszewski, T.~Paterek, and T.~C.~H. Liew,
		``Quantum reservoir processing,''
		\textit{npj Quantum Information} \textbf{5}, 35 (2019).
		\bibitem{MODEL5}
		B.~L.~Fang, J.~Shi, and T.~Wu,
		``Quantum-memory-assisted entropic uncertainty relation and quantum coherence in structured reservoir,''
		\textit{Int. J. Theor. Phys.} \textbf{59}, 763 (2020).
		\bibitem{Oularabi2025}
		A. Oularabi, A. El Allati, and K. El Anouz, ``Enhancing ergotropy of quantum batteries through coherence and non-markovianity,'' \textit{Physica A: Statistical Mechanics and its Applications}, 131003 (2025).
		\bibitem{Aiache2025}
		Y. Aiache, A. Ullah, Ö. E. Müstecaplıoğlu, and A. El Allati, ``Quantum metrology of a structured reservoir,'' \textit{Phys. Rev. A} \textbf{111}, 062619 (2025).
		
		\bibitem{MODEL6}
		N.~Linden, S.~Popescu, and P.~Skrzypczyk,
		``How small can thermal machines be? The smallest possible refrigerator,''
		\textit{Phys. Rev. Lett.} \textbf{105}, 130401 (2010).
		\bibitem{MODEL7}
		A.~Khoudiri, A.~El~Allati, \"{O}.~E.~M\"{u}stecapl\i o\u{g}lu, and K.~El~Anouz,
		``Non-Markovianity and a generalized Landauer bound for a minimal quantum autonomous thermal machine with a work qubit,''
		\textit{Phys. Rev. E} \textbf{111}, 044124 (2025).
		
		\bibitem{MODEL4a}
		D.~C.~McKay, S.~Filipp, A.~Mezzacapo, E.~Magesan, J.~M.~Chow, and J.~M.~Gambetta,
		``Universal gate for fixed-frequency qubits via a tunable bus,''
		\textit{Physical Review Applied} \textbf{6}, 064007 (2016).
		
		\bibitem{MODEL5a}
		A.~P.~Place, L.~V.~Rodgers, P.~Mundada, B.~M.~Smitham, M.~Fitzpatrick, Z.~Leng, \textit{et al.},
		``New material platform for superconducting transmon qubits with coherence times exceeding 0.3 milliseconds,''
		\textit{Nature Communications} \textbf{12}, 1779 (2021).
		
		\bibitem{MODEL6a}
		
		A. Khoudiri, K. El Anouz, A. El Allati,
		``Generation of Quantum Entanglement in Autonomous Thermal Machines: Effects of Non‑Markovianity, Hilbert Space Structure, and Quantum Coherence,''
		\textit{arXiv:2508.18056v2} [quant‑ph] (2025).
		
		
		\bibitem{MODEL7a}
		M.~Müller, K.~Hammerer, Y.~L.~Zhou, C.~F.~Roos, and P.~Zoller,
		``Simulating open quantum systems: from many-body interactions to stabilizer pumping,''
		\textit{New Journal of Physics} \textbf{13}, 085007 (2011).
		
		
		\bibitem{MODEL8a}
		C.~Xiong, H.~Li, H.~Xu, M.~Zhao, B.~Zhang, C.~Liu, and K.~Wu,
		``Coupling effects in single-mode and multimode resonator-coupled system,''
		\textit{Optics Express} \textbf{27}, 17718--17728 (2019).
		
		\bibitem{MODEL9a}
		J.~Majer, J.~M.~Chow, J.~M.~Gambetta, J.~Koch, B.~R.~Johnson, J.~A.~Schreier, \textit{et al.},
		``Coupling superconducting qubits via a cavity bus,''
		\textit{Nature} \textbf{449}, 443--447 (2007).
		
		\bibitem{MODEL11}
		M.~Lostaglio, D.~Jennings, and T.~Rudolph,
		``Description of quantum coherence in thermodynamic processes requires constraints beyond free energy,''
		\textit{Nat. Commun.} \textbf{6}, 6383 (2015).
		
		\bibitem{MODEL14}
		E.~Chitambar and G.~Gour,
		``Comparison of incoherent operations and measures of coherence,''
		\textit{Phys. Rev. A} \textbf{94}, 052336 (2016).
		\bibitem{MODEL15}
		W.~De~Roeck and C.~Maes,
		``Quantum version of free-energy–irreversible-work relations,''
		\textit{Phys. Rev. E} \textbf{69}, 026115 (2004).
		\bibitem{MODEL16}
		T.~Baumgratz, M.~Cramer, and M.~B.~Plenio,
		``Quantifying coherence,''
		\textit{Phys. Rev. Lett.} \textbf{113}, 140401 (2014).
		\bibitem{MODEL17}
		J.~Majer, J.~M.~Chow, J.~M.~Gambetta, J.~Koch, B.~R.~Johnson, J.~A.~Schreier, \textit{et al.},
		``Coupling superconducting qubits via a cavity bus,''
		\textit{Nature} \textbf{449}, 443--447 (2007).
		\bibitem{MODEL18}
		L.~DiCarlo, J.~M.~Chow, J.~M.~Gambetta, L.~S.~Bishop, B.~R.~Johnson, D.~I.~Schuster, \textit{et al.},
		``Demonstration of two-qubit algorithms with a superconducting quantum processor,''
		\textit{Nature} \textbf{460}, 240--244 (2009).
		\bibitem{MODEL19}
		L.~DiCarlo, J.~M.~Chow, J.~M.~Gambetta, L.~S.~Bishop, B.~R.~Johnson, D.~I.~Schuster, \textit{et al.},
		``Demonstration of two-qubit algorithms with a superconducting quantum processor,''
		\textit{Nature} \textbf{460}, 240--244 (2009).
		\bibitem{MODEL20}
		D.~C.~McKay, S.~Filipp, A.~Mezzacapo, E.~Magesan, J.~M.~Chow, and J.~M.~Gambetta,
		``Universal gate for fixed-frequency qubits via a tunable bus,''
		\textit{Phys. Rev. Appl.} \textbf{6}, 064007 (2016).
		\bibitem{MMODL21}
		S.~Zakavati, F.~T.~Tabesh, and S.~Salimi,
		``Bounds on charging power of open quantum batteries,''
		\textit{Phys. Rev. E} \textbf{104}, 054117 (2021).
		\bibitem{Model25}
		D.~Murphy, A.~Kiely, I.~D'Amico, and S.~Campbell,
		``Ergotropy transport in a one-dimensional spin chain,''
		\textit{Physical Review A} \textbf{112}(5), 052214 (2025).
		\bibitem{Model26}
		M.~J.~Sarmah and H.~P.~Goswami,
		``Noise-induced coherent ergotropy of a quantum heat engine,''
		\textit{Physical Review A} \textbf{110}(3), 032213 (2024).
				\bibitem{MODEL17_correct}
		G. L. Giorgi and S. Campbell, “Multipartite quantum and classical correlations in symmetric n-qubit mixed states,” Quantum Inf. Process. 15, 4599–4611 (2016), DOI: 10.1007/s11128-016-1421-x.
		\bibitem{Model27}
		J.-Y.~Gyhm and U.~R.~Fischer,
		``Beneficial and detrimental entanglement for quantum battery charging,''
		\textit{AVS Quantum Sci.} \textbf{6}, 12001 (2024).
		\bibitem{FINAL1}
		G.~Francica,
		``Quantum correlations and ergotropy,''
		\textit{Phys. Rev. E} \textbf{105}, L052101 (2022).
		\bibitem{FINAL2}
		J.~P.~Santos, L.~C.~Céleri, G.~T.~Landi, and M.~Paternostro,
		``The role of quantum coherence in non-equilibrium entropy production,''
		\textit{npj Quantum Inf.} \textbf{5}, 23 (2019).
		\bibitem{FINAL3}
		D.~Šafránek, D.~Rosa, and F.~C.~Binder,
		``Work extraction from unknown quantum sources,''
		\textit{Phys. Rev. Lett.} \textbf{130}, 210401 (2023).
		\bibitem{FINALE}
		G.~Francica, G.~Binder, G.~Guarnieri, M.~T.~Mitchison, J.~Goold, and F.~Plastina,
		``Quantum coherence and ergotropy,''
		\textit{Phys. Rev. Lett.} \textbf{125}, 180603 (2020).
		
		\bibitem{FINAL1}
		P.~Skrzypczyk, R.~Silva, and N.~Brunner,
		``Passivity, complete passivity, and virtual temperatures,''
		\textit{Phys. Rev. E} \textbf{91}, 052133 (2015).

	\end{thebibliography}
\end{document}